\documentclass[twocolumn]{aastex631}

\usepackage[utf8]{inputenc}
\usepackage{amsmath}
\usepackage{ulem}
\usepackage{bm}
\usepackage[caption=false]{subfig}

\shorttitle{AASTeX v6.3.1 Sample article}
\shortauthors{Narloch et al.}
\graphicspath{{./}{figures/}}

\begin{document}

\title{Period-Luminosity Relations for Galactic classical Cepheids 
in the Sloan bands\footnote{Based on data from the Las Cumbres Observatory}}

\author[0000-0003-2335-2060]{Weronika Narloch}
\affiliation{Universidad de Concepci\'on, Departamento de Astronomia, Casilla 160-C, Concepci\'on, Chile} 
\affiliation{Nicolaus Copernicus Astronomical Center, Polish Academy of Sciences, Bartycka 18, 
00-716 Warszawa, Poland} 

\author{Gergely Hajdu}
\affiliation{Nicolaus Copernicus Astronomical Center, Polish Academy of Sciences, Bartycka 18, 
00-716 Warszawa, Poland} 


\author{Grzegorz Pietrzy\'nski}
\affiliation{Universidad de Concepci\'on, Departamento de Astronomia, Casilla 160-C, Concepci\'on, Chile} 
\affiliation{Nicolaus Copernicus Astronomical Center, Polish Academy of Sciences, Bartycka 18, 
00-716 Warszawa, Poland}

\author{Wolfgang Gieren}
\affiliation{Universidad de Concepci\'on, Departamento de Astronomia, Casilla 160-C, Concepci\'on, Chile} 

\author{Piotr Wielg\'orski}
\affiliation{Nicolaus Copernicus Astronomical Center, Polish Academy of Sciences, Bartycka 18, 
00-716 Warszawa, Poland} 

\author{Bart\l omiej Zgirski}
\affiliation{Nicolaus Copernicus Astronomical Center, Polish Academy of Sciences, Bartycka 18, 
00-716 Warszawa, Poland} 

\author{Paulina Karczmarek}
\affiliation{Universidad de Concepci\'on, Departamento de Astronomia, Casilla 160-C, Concepci\'on, Chile} 

\author{Marek G\'orski}
\affiliation{Nicolaus Copernicus Astronomical Center, Polish Academy of Sciences, Bartycka 18, 
00-716 Warszawa, Poland} 

\author{Dariusz Graczyk}
\affiliation{Nicolaus Copernicus Astronomical Center, Polish Academy of Sciences, Rabia\'nska 8, 87-100 
Toru\'n, Poland} 



\begin{abstract}

We present the first period--luminosity (PL) and period--Wesenheit (PW) 
relations in the Sloan--Pans-STARRS $g_{P1}r_{P1}i_{P1}$ bands for 
classical fundamental mode Cepheids in the Milky Way. We used a~relatively 
modest number of $76$ stars for the PL and $84-85$ stars for the PW relations 
calibration. The data for the project were collected with the network of 
40-cm telescopes of Las Cumbres Observatory, and Gaia Data Release 3 parallaxes 
were used for the calculations. 
These $gri$-band PL and PW  relations calibrations will be a~useful tool for 
distance determinations in the era of large sky surveys using the Sloan photometric 
system, especially with the near-future start of the Large Synoptic Survey of 
Space and Time (LSST).

\end{abstract}



\section{Introduction} \label{sec:intro}

Since the discovery of a~relationship connecting the periods and luminosities of 
classical Cepheids in the Small Magellanic Cloud (SMC) by Henrietta Swan Leavitt 
\citep{Leavitt1908,LP1912}, the period-luminosity (PL) relation (or Leavitt law) 
became one of the most important methods for estimating distances in the universe.
The law serves as the first rung of the extragalactic distance ladder and is used 
to measure the distances to type Ia supernova host galaxies, which is a~crucial 
step in the determination of the Hubble constant \citep[e.g.,][]{Riess2022}. 
For this reason classical Cepheids, together with other types of pulsating stars, 
such as type II Cepheids and RR Lyrae stars, became very important distance 
indicators.  
Astronomers dedicate considerable effort establishing ever more precise and accurate 
PL relations, as well as period--Wesenheit (PW) relations, where the Wesenheit index 
is a~reddening-free magnitude by construction \citep{Madore1982}. 

PL and PW relations for classical Cepheids are available for a~wide range of optical 
\citep[e.g.,][]{Udalski1999,Tammann2003,Fouque2007} and near- and mid-infrared bands 
\citep[e.g.,][]{Macri2015,Ripepi2016,Wang2018}, which have recently been extensively 
studied for their dependence on metallicity 
\cite[e.g.,][]{Gieren2018,Ripepi2020,Breuval2021,Trentin2023,Molinaro2023} and 
also theoretically \cite[e.g.,][]{Anderson2016,DeSomma2022}. 
Near-infrared PL relations have undeniable advantage over the optical range, mainly 
due to their decreased sensitivity to extinction, producing much smaller scatter. 
Nonetheless, the more symmetric shapes and smaller amplitudes of Cepheids at longer 
wavelengths, can cause the problem with their identification in the near-infrared 
domain \citep{PSU2021}. 
This means that the optical range still plays an essential role for the discovery 
and classification of new variable stars, and precise PL relations are intensively 
studied for many optical filters. 
However, to our knowledge, the Sloan filters have not been widely used in the context 
of Cepheids so far, except for studies of variables in other galaxies, e.g., 
\citet{HM2015} for NGC4258 or \citet{AL2023} for M33.  

The Sloan photometric system \citep[$ugriz$;][]{Fukugita1996} is a~wide-band photometric 
system having wavelength coverage from $3000$ up to $11000\,\rm{\AA}$ and was developed 
for the Sloan Digital Sky Survey \citep[SDSS;][]{Abazajian2003}. 
Since its inception, this photometric system has gained wide-spread popularity for 
large-scale sky surveys. The most noteworthy are the Panoramic Survey Telescope And 
Rapid Response System \citep[Pan-STARSS;][]{Tonry2012}, where the first part of the 
data release (Pan-STARRS1) of Andromeda (PAndromeda) provided PL relationships for 
pulsating stars in that galaxy \citep{Kodric2018}, and the Dark Energy Survey, using 
the DECam wide-field camera \citep{Flaugher2015,DES2016}.  
The Pan-STARSS and the Dark Energy Survey have become {\it de facto} standards 
for imaging in the Sloan bands for the northern and southern hemispheres, respectively.

Another ambitious upcoming project is the $10$-year Legacy Survey of Space and Time 
\citep[LSST;][]{Ivezic2019}, which will be carried out at the Vera C. Rubin Observatory, 
currently under construction. The LSST is expected to provide an enormous amount of data 
and discover many new variable stars, among others, Cepheids in the Local Group galaxies 
and beyond. Hence, new possibilities are opening up in the field of distance measurements. 
However, in order to take full advantage of this project, precise calibration of PL 
relations for classical Cepheids (and other type of pulsating stars) has to be performed 
in the Sloan passbands, particularly in the Milky Way (MW), where good-quality geometric 
parallaxes for such objects are provided by the \textit{Gaia} mission \citep{GaiaCol2023}. 

It is important to note, that unlike other photometric systems used before, generally, 
wide-field surveys in the Sloan system are not transforming their observations to the 
original reference system defined by \citet[][where the magnitudes are often denoted as 
$u'g'r'i'z'$]{Smith2002}, which is defined as the native system of the 1-m telescope of 
the US Naval Observatory Flagstaff Station. Instead, each survey releases their photometry 
in their own native systems, where the zero-points of the bands are calibrated using 
the transformed magnitudes from \citet{Smith2002}.
Therefore, it is a crucial detail to clarify exactly which system a given study is using. 


Realizing the need for well-calibrated PL relationships in the Sloan bands, various PL 
and Wesenheit relationships have been established for different kinds of pulsating variable 
stars using data from the Zwicky Transient Facility 
\citep[ZTF;][]{Bellm2018,Bellm2019,Dekany2020}. These include RR Lyrae variables 
\citep{Ngeow2022_RRL}, type II \citep{Ngeow2022_T2CEP} and anomalous Cepheids  
\citep{Ngeow2022_AC}, and even SX Phe variables \citep{Ngeow2023_SXPhe}. However, 
due to the relatively low number of calibrators for type II and anomalous Cepheids, as 
well as problems with blending due to the usage of stars in globular clusters and the 
large pixel size of the ZTF camera, significant systematic uncertainties remain in these 
calibrations. Furthermore, there is still no such calibration available for classical
Cepheids, given that they are generally not located in clusters, and the vast majority 
of the ones with reliable Gaia parallaxes are brighter than the ZTF saturation limit.

The aim of this work is to derive the PL and PW relations for MW classical Cepheids 
in the three Sloan $gri$ filters, and specifically calibrated to the Pan-STARRS system, 
for the first time. We expect that the relationships obtained here will become a~useful 
tool for measuring distances in the universe in the upcoming era of large programs like 
the LSST. 

The paper is organized as follows. In Sect.~\ref{sec:data} we describe the 
sample of the selected classical Cepheids, the reduction of the data and further 
selection criteria for the derivation of the PL and PW relations, described later in 
Sect.~\ref{sec:relations}. A~short discussion is given in Sect.~\ref{sec:discussion}, 
and we conclude with a~summary of our main results in Sect.~\ref{sec:summary}. 


\section{Data} \label{sec:data}

\subsection{Sample of stars} \label{ssec:sample}

\begin{figure*}[ht!]
\plotone{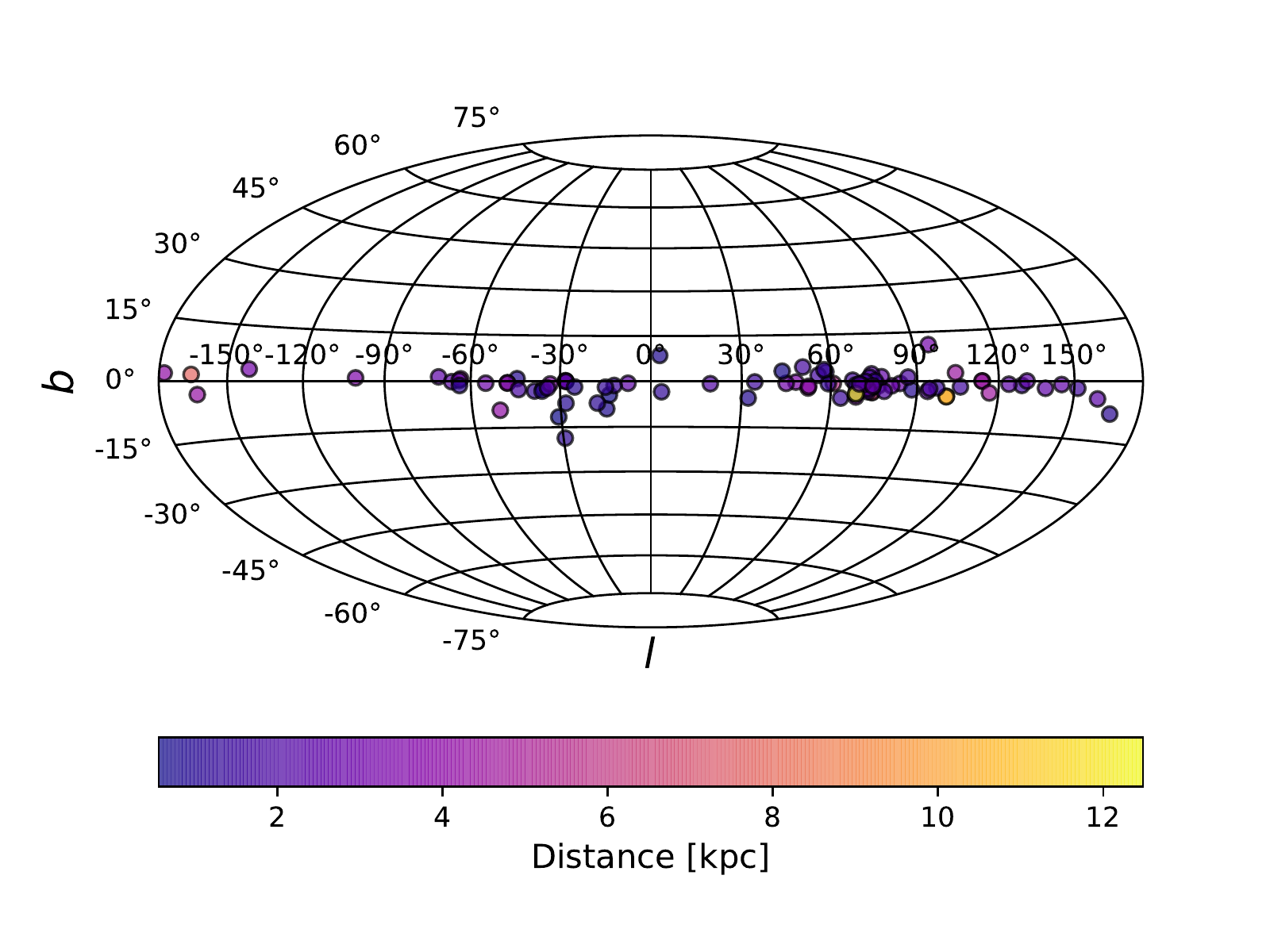}
\caption{Location of the classical Cepheids used for establishing the PL relations 
in this paper given in Galactic coordinates. \label{fig:galmap}}
\end{figure*}

For the purpose of deriving the PL relations we observed $96$ classical Cepheids 
pulsating in the fundamental mode. The chosen stars are bright, with the great 
majority having Gaia $G$-magnitudes in the range from about $6$ to $12$~mag 
with a~few fainter stars up to $15.7$~mag, which were mostly observed serendipitously, 
as they lie in the same field as a~bright (primary) target.
They are located at distances between $0.57$ and $12.48$~kpc (with a~median 
distance of $2.3$~kpc), which assure good-quality Gaia DR3 parallaxes \citep{GaiaCol2023}. 
Their periods taken from \citet{PSU2021}, ranging from about 
$3$ to $69$~days. 
The locations of the selected Cepheids are marked in Fig.~\ref{fig:galmap} and 
their main physical parameters are summarized in Table~\ref{tab:cepheids}. 

\subsection{Data and reduction} \label{ssec:data_reduction}

The data were collected between May $2020$ and July $2022$ within various observing 
programs\footnote{CLN2020A-003, CLN2020B-008, CLN2021A-009, and CLN2022A-008}. 
The images of target stars were obtained in three filters modeled after the 
original Sloan $g'r'i'$ filters, using $16$ robotic 40-cm telescopes, which are 
part of the Las Cumbres Observatory (LCO) Global Telescope 
Network\footnote{\url{https://lco.global/}}. 
They are equipped with $3\rm{K} \times 2\rm{K}$ SBIG STL-6303 cameras, with 
a~field of view of $29.2 \times 19.5$~arcmin$^2$ and pixel size of $0.571$~arcsec 
pixel$^{-1}$ without binning. Observations were taken in the air mass range 
of $1.00-1.87$, and measured average seeing of about $2.35$, $2.26$ and 
$2.30$~arcsec in the Sloan $g$, $r$ and $i$ bands, respectively.  
We aimed to obtain well-covered light curves for each of our targets. Therefore, 
during the course of the project, their observations were continuously monitored, 
and new observing requests were sent to LCO to fill gaps in the light curve phases.

The raw images were reduced and processed with the LCO BANZAI 
pipeline\footnote{\url{https://lco.global/documentation/data/BANZAIpipeline/}}
and we accessed them from the LCO Archive\footnote{\url{https://archive.lco.global/}}.
We performed aperture photometry on these images with a~fixed aperture of $14$~pixels, 
corresponding to about $8$~arcsec (which for the vast majority of images contain 
virtually all light from  the stars) using the standard DAOPHOT package \citep{Stetson1987}.
Next, we cross-matched the list of stars from each image with the ATLAS All-Sky 
Stellar Reference Catalog version 2 \citep[ATLAS-REFCAT2;][]{Tonry2018}, which 
provides a compilation of magnitudes for stars between magnitudes $6$ and $19$ in 
the Pan-STARRS system (for the bands used here, generally denoted as $g_{P1}r_{P1}i_{P1}$).
The average DAOPHOT photometric errors were $0.04$~mag for Sloan $gr$ and $0.05$~mag 
in Sloan $i$ for stars with Sloan $g < 14.0$~mag, which are the magnitudes of most 
of our local reference stars. The average DAOPHOT photometric errors for the Cepheids 
themselves are $0.002$~mag in all three filters. 

In order to obtain the light curves of the Cepheids in our fields, first we identified 
any possible variable stars around our target Cepheids, and removed them from the 
list of potential comparison stars. 
From the remaining ones, we selected stars with photometric uncertainties better than 
$0.1$~mag and up to a~few magnitudes fainter than the target Cepheid in the field. We 
then accepted them as the reference points for standardization that we used to calculate 
the coefficients in the following transformation equations: 

\begin{eqnarray} 
\label{eq:tran1}
g_{inst} = a_1 \cdot g_{cat} + a_2 + a_3 \cdot (g-r)_{cat} \\ 
\label{eq:tran2}
r_{inst} = b_1 \cdot r_{cat} + b_2 + b_3 \cdot (g-r)_{cat} \\  
\label{eq:tran3}
i_{inst} = c_1 \cdot i_{cat} + c_2 + c_3 \cdot (g-i)_{cat},
\end{eqnarray}

\noindent where $g_{inst}$, $r_{inst}$ and $i_{inst}$ are our Sloan instrumental 
magnitudes normalized to one second, while $g_{cat}$, $r_{cat}$ and $i_{cat}$ are the 
catalog magnitudes from \citet{Tonry2018} in the Pan-STARRS system. 
The colors in the equations were calculated based on three subsequent images in each 
filter, which was possible as the observation strategy was to perform consecutive exposures 
in all three bands for all epochs. 
We calculated the $a_3,\,b_3$, and $\,c_3$ coefficients based on the constant-magnitude stars 
from all fields for a~given telescope, and then fixed them for that telescope, in order to 
reduce the number of derived coefficients. 
The $a_2,\,b_2$, and $\,c_2$ coefficients are the zero-points of the transformations. 
The $a_1,\,b_1$, and $\,c_1$ coefficients were added in order to remove the trend that appeared 
in the residuals when they were not present in the equations. 
The reason for the appearance of this trend is the presence of nonlinearity affecting 
the SBIG~6303 cameras, as noted before by the LCO 
staff\footnote{\url{https://lco.global/documentation/}}. 
While a~procedure was proposed to minimize it, the authors of the report pointed out 
themselves that the form of the proposed correction is not optimal. 
We noticed that adding the $a_1,\,b_1$, and $\,c_1$ coefficients in our transformation equations 
removed the trend well, so we decided to stick to this solution. The number of used 
comparison stars in a~given image ranged from a few to several dozens. The fitting procedure 
was done iteratively by removing deviating stars in each of three iterations (first by 
using the Random Sample Consensus algorithm from the {\tt scikit-learn} Python package, and 
then two $3\sigma$~clipping steps). In the end, the residuals were inspected by eye, and single 
outliers were discarded if necessary. 
The derived coefficients from Equations~(\ref{eq:tran1})-(\ref{eq:tran3}) were then used 
to transform the brightnesses of the Cepheids in a~given frame to the Pan-STARSS photometric 
system. The typical rms value of the transformations was about $0.035$~mag, but depended 
on the quality of images and the number of comparison stars used. 
We also performed a~bootstraping procedure with $1000$ iterations for 
Equations~(\ref{eq:tran1})-(\ref{eq:tran3}) to estimate the errors of the standardized magnitudes 
of the Cepheids in each frame, which we utilized in the next step. 
The mean shift between our calibrated and catalog magnitudes in the range of our Cepheid 
magnitudes in the three Sloan filters is less than $0.01$~mag for all but one camera. 
Thus, this value can be adopted as the maximum error of the accuracy of the photometric 
zero-point, which propagates to the final PL/PW relations as a~systematic error. 

For each star in each of the three Sloan filters we phased the data, and subsequently 
calculated intensity-averaged mean apparent magnitudes by fitting the Fourier series to 
the Cepheid light curve with the {\tt curve\_fit} Python algorithm. The applied Fourier 
order depended on the properties of a~given light curve and was chosen by visual inspection. 
The minimum chosen order was equal to two and the maximum used was $14$. High orders were 
allowed to fit well the maximum brightness in specific cases. 
During the fitting procedure we applied weights to each point, which we calculated 
from the errors determined from bootstraping in the previous step. 
Additionally, we performed Monte Carlo simulations, but errors estimated with this method
were of an order smaller than the ones returned from Fourier series fit, so we conservatively 
adopted the latter as the uncertenties of obtained mean apparent magnitudes. 
The final light curves with fitted Fourier series are presented in Fig.~\ref{fig:fig1}. 

\subsection{Reddening} \label{ssec:redd}

The reddening of Cepheids is a~complicated topic and variously treated in 
the literature (e.g., \citealp{Harris1981}; \citealp{Turner1989}; \citealp{Kovtyukh2008}; 
\citealp[see][and references therein for a~review]{Turner2016}; \citealp{MFM2017}).
Multiple extinction maps have been published for our Galaxy, e.g., 
\citet[][hereafter SFD map]{SFD} or the more recent Gaia/2MASS 3D Dust Extinction Map 
\citep{Lallement2019} and {\tt Bayestar2019} reddening map \citep{Green2019} available 
through the {\tt dustmap} code\footnote{\url{https://dustmaps.readthedocs.io/en/latest/}}, 
all having their advantages and disadvantages. For the Cepheids in our sample, 
however, those maps turned out to be inadequate. Even though the SFD map provides 
reddening values for all our target stars, it turned out to be overestimated in most 
of the cases even after accounting for the simple model of dust in the Galaxy in the 
line of sight.
As for the Gaia/2MASS 3D map, it is not available for about a~half of our stars, as 
they are located at distances larger than $3$~kpc which is its limit. Similarly, the 
{\tt Bayestar2019} reddening map did not provide reddening values for about a~half of 
the objects, due to a~lack of coverage of specific Galactic regions. Despite the 
abovementioned problems, we have tested the maps but they generated an artificially large 
scatter in the resulting PL relations.
For this reason we have decided to follow the approach from \citet{Breuval2021} and 
adopt reddening values from 
\citet{Fernie1995}\footnote{\url{https://www.astro.utoronto.ca/DDO/research/cepheids/table_colourexcess.html}}, 
after application of the $0.94$ scaling factor suggested by \citet{Groenewegen2018}.  
Also, we have adopted an uncertainty of $0.05$~mag if not provided. 
Reddening information was only available for $86$ stars from our sample (see column~6 
in Table~\ref{tab:cepheids}), but as we still had far enough objects for the construction 
of precise PL relations, we decided against the idea of mixing reddenings from different 
sources to keep consistency.  

And finally, we adopted extinction vectors ($R_{\lambda}$) for the Pan-STARRS1 
$g_{P1}r_{P1}i_{P1}$ filters from \citet[][their Table~1]{Green2019}, equaling: 
$R_{g} = 3.518$, $R_{r} = 2.617$ and $R_{i} = 1.971$.   
We also calculated $R_{\lambda}$ directly from the \citet{F99} reddening law using 
the {\tt extinction}\footnote{\url{https://extinction.readthedocs.io/en/latest/}} 
Python package, assuming $R_{V} = 3.1$, and effective wavelengths for the Pan-STARSS 
filters from Table~4 of \citet{Tonry2012}, and obtained  
$R_{g} = 3.666$, $R_{r} = 2.583$ and $R_{i} = 1.901$.


\subsection{Distances} \label{ssec:dist} 

In order to calibrate PL relations, one needs to recalculate the derived apparent 
mean magnitudes of stars into their absolute magnitudes, which requires knowing the 
distances to the stars. 
The best source of geometric parallaxes known today is the third Gaia Data Release 
(hereafter Gaia DR3) catalog \citep{GaiaCol2023}. 
\citet{Lindegren2021a} proposed a~correction for the zero-point (ZP) offset of Gaia 
parallaxes, which can be calculated with the dedicated Python 
code\footnote{\url{https://gitlab.com/icc-ub/public/gaiadr3_zeropoint}} they provided. 
This correction takes into account the ecliptic latitude, magnitude, and color of a~given 
star. The parallaxes of our Cepheids range between about $0.08$ to $1.76$~mas (see 
column~3 in Table~\ref{tab:cepheids}), and the corrections from about $-9$ to $-52\,\mu$as 
(with a~mean of about $-29\,\mu\rm{as}$). 
Following the recommendation of \citet{Lindegren2021a} to include an uncertainty of a~few 
$\mu$as in the ZP, we adopted $5\,\mu$as as a~systematic uncertainty. This value is 
propagated into the distance modulus, and hence the PL relation determination, as 
$0.025$~mag for the median parallax for our sample, equal to $0.43$~mas. 

The quality of the parallaxes can be judged based on two parameters: the Renormalized 
Unit Weight Error (RUWE) and the Goodness-of-fit (GOF). 
The RUWE indicator is sensitive to the photocentric motions of unresolved objects, such 
as astrometric binaries. Ideally, it should be equal to $1.0$ \citep{Lindegren2021a}. 
The GOF parameter was used by, e.g., \citet{Riess2021} as an indicator of the level of 
asymmetry of a~source, and was adopted by the authors to be less than $12.5$. 
Following \citet{Breuval2021} and \citet{Wielgorski2022} we rejected stars with 
$\rm{RUWE}>1.4$ (seven stars) or $\rm{GOF}>12.5$ (five stars) from our sample of Cepheids. 
All stars that met the second condition also met the first, which means that RUWE is more 
constraining than GOF in our case. The rejected stars were FN~Vel, RZ~CMa, SU~Cru, U~Aql, 
UW~Car, V496~Aql, and WW~Car (marked in Table~\ref{tab:cepheids} with the "a" footnote in 
their IDs). Moreover, we have rejected from the fitting two additional stars (BB~Sgr and 
IT~Car) because of the poor quality of their light curves (marked in Table~\ref{tab:cepheids} 
with the "b" footnote, see also Fig.~\ref{fig:fig1}). 

Five of our Cepheids fall into the problematic range of transition of Gaia window classes 
where the parallax ZP could be affected \citep[see Fig.~1 in][]{Lindegren2021b}. 
The three of them fall into $G=11.0 \pm 0.2$~mag and two into $G=12.0 \pm 0.2$~mag. 
Following \citet{Breuval2021}, we quadratically added $10\,\mu$as to the parallax uncertainties  
of the affected stars. Finally, following the suggestion of \citet{Riess2021} we increased 
all Gaia DR3 parallax uncertainties by $10\%$ to account for a~possible excess uncertainty. 
The mean of the calculated parallax uncertainties (given in column~3 in Table~\ref{tab:cepheids})  
is $21\,\mu$as.

\section{Derived Relations} \label{sec:relations}

\subsection{Period-Luminosity Relations} \label{ssec:plr} 

To calibrate the PL relations for a~sample of selected classical MW Cepheids in the 
three Sloan filters, $g_{P1}r_{P1}i_{P1}$, first we dereddened the apparent mean magnitudes 
($m_{\lambda}$) derived from the Fourier series fitting as described in Sec.~\ref{ssec:redd}, 
and then calculated absolute magnitude ($M_{\lambda}$) of each Cepheid as: 

\begin{equation} \label{eq:absmag}
 M_{\lambda} = m_{\lambda} + 5\log_{10} \varpi + 5,
\end{equation}

\noindent where $\varpi$ is the parallax in arcseconds. Next, we plotted the derived absolute 
magnitudes against the decimal logarithm of the periods related by the following relationship: 

\begin{equation} \label{eq:plr}
 M_{\lambda} = a_{\lambda} (\log P - \log P_0) + b_{\lambda},
\end{equation}

\noindent where $a_{\lambda}$ and $b_{\lambda}$ are the slope and intercept we are looking for, 
respectively. 
The pivot period $logP_0 = 1$ was added to the original equation in order to minimize 
correlation between the slope and the intercept. It was adopted to be equal to $1$, because 
we wanted to use the value close to the median of our Cepheid sample ($\sim 0.98$~days) 
but simultaneously being a~round number. 

We used the {\tt curve\_fit} function from the {\tt scipy} Python library for the fitting 
in each passband with $3\sigma$-clipping and obtained the slope and intercept from 
Equation~(\ref{eq:plr}) together with their corresponding errors, which we adopted as the final 
statistical uncertainties of the calculated coefficients. The obtained values are given in 
Table~\ref{tab:plr}, where $76$ stars were used for the final fit from the range of distances 
from $0.6$ to $7.8$~kpc (with a~median of about $2.2$~kpc).
As a~control test, we also performed a~bootstraping procedure on the residuals with $10,000$ 
iterations in each passband and got very similar values for the uncertainties compared to 
the ones from the least-square method, which assures us that the adopted slope and intercept 
errors are reliable. 
The resulting rms of the PL relation is $0.25$, $0.23$, and $0.22$ in the Sloan -- Pan-STARRS 
$g_{P1}r_{P1}i_{P1}$ bands, respectively, for \citet{Green2019} as the source of reddening 
vectors, and analogously $0.25$, $0.22$, and $0.22$ for the \citet{F99} vectors.  

To avoid any bias in the absolute magnitude due to its nonlinear relationship with parallax, 
we also calculated the Astrometry-Based Luminosity (ABL), following the approach recommended 
by \citet{FC1997} and \citet{AL1999}, where: 

\begin{equation} \label{eq:abl}
 ABL_{\lambda} = 10^{0.2M_{\lambda}} = \varpi_{(arcsec)} 10^{\frac{m_{\lambda + 5}}{5}},
\end{equation}

\noindent Then the PL relation given by Equation~(\ref{eq:plr}) will take the form: 

\begin{equation} \label{eq:abl_plr}
 ABL_{\lambda} = 10^{0.2[a_{\lambda}(logP - logP_0) + b_{\lambda}]}.
\end{equation}

We performed the fitting of Equation~(\ref{eq:abl_plr}) as described above and the resulting 
coefficients with their corresponding uncertainties are given in Table~\ref{tab:plr}. 

The PL relations for the three filters are presented in Fig.~\ref{fig:plr}. 
The error bars in the plot are derived from the error propagation, taking into account 
the statistical errors on the mean magnitudes and parallaxes, where the latter ones are dominant. 
Solid and dashed lines represent the fit from the classical and ABL methods, respectively. 
The coefficients from these two approaches differ slightly, and the difference is higher 
for slopes (within $2\sigma$), while the intercepts are within $1\sigma$. 



\subsection{Period-Wesenheit Relations} \label{ssec:pwr}

PL relations based on reddening-free Wesenheit indices \citep[or sometimes called Wesenheit 
magnitudes, originally introduced by][]{Madore1982} have proved to be particularly useful 
for distance determination due to their independence of reddening.
Following \citet{Ngeow2022_T2CEP} we defined three Wesenheit indices, 
$W_{r}^{ri}$, $W_{r}^{gr}$, and $W_{g}^{gi}$, for which, however, we calculated coefficients 
based on $R_{\lambda}$ from two sources: \citet{Green2019} and 
\citet[][]{F99}; see Section~\ref{ssec:redd}. 
The resulting equations are given in column~2 of Table~\ref{tab:pwr}. 
The mean apparent magnitudes used for deriving the Wesenheit indices were not corrected for 
reddening. We then fitted the relations analogous to Equation~(\ref{eq:plr}) and Equation~(\ref{eq:abl_plr}): 

\begin{equation}
 W = a (\log P - \log P_0) + b \label{eq:pwr},
\end{equation}
\begin{equation}
 ABL_{W} = 10^{0.2[a(\log P - \log P_0) + b]} \label{eq:abl_pwr},
\end{equation}


\noindent where again $a$ and $b$ are the slope and intercept sought, respectively, and 
$\rm{log}P_0$ is the pivot period, chosen to be $1$, and we applied the $3\sigma$-clipping 
procedure to remove outliers. 
Because PW relations do not require reddening values, the number of stars used for 
calibration was higher than for the case of absolute magnitudes, and ranged between $84$ 
and $85$ stars (from a~distance range between $0.6$ and $12.5$~kpc, with a~median of $2.4$~kpc).
The smallest rms we found was for the $R_{\lambda}$ calculated based on 
\citet[][$\rm{rms} = 0.21 - 0.27$]{F99} and the largest was based on  
\citet[][$\rm{rms} = 0.27 - 0.30$]{Green2019}.   
We performed the fitting for the Wesenheit coefficient with the ABL method, as well. 
The coefficients with their corresponding uncertainties returned by the least-squares 
algorithm are given in Table~\ref{tab:pwr}. 
Figure~\ref{fig:pwr} presents an~example of the fit for parameter $R_{\lambda}$ from the 
Wesenheit index calculated based on the extinction coefficients from \citet{Green2019}. 
Also in this case the slope values from the classical and ABL methods differ noticeably, 
while the intercepts agree very well within the errors. 

%

\section{Discussion} \label{sec:discussion}

\subsection{Stars rejected from the PL/PW relation fit} \label{ssec:rejectedstars}

As described in Section~\ref{ssec:dist}, seven stars from the original sample were rejected 
from the fitting because they have not met the criteria of the Gaia RUWE and GOF quality 
indicators. We marked these stars in Figures~\ref{fig:plr} and \ref{fig:pwr}. 
Six of them (FN~Vel, RZ~CMa, U~Aql, UW~Car, V496~Aql, and WW~Car) lie very close to the 
found relations, but the obvious outlier is SU~Cru, which deviates significantly toward 
brighter absolute magnitudes. This variable is defined in the Simbad Astronomical 
Database\footnote{\url{http://simbad.cds.unistra.fr/simbad/}} as a~double or multiple star. 
\citet{Szabados1996} listed this star as one of the spectroscopic binaries found among 
classical Cepheids, which has a~red photometric companion. 
Additionally, we have also rejected BB~Sgr and IT~Car because of the bad quality of their 
light curves (see Figure~\ref{fig:fig1}).

After $3\sigma$-clipping, AD~Cam was rejected from the fit of the PL relations for being too 
faint. It is also deviating in the PW relations, although not so significantly. The light 
curve of that star presented in Figure~\ref{fig:fig1} is not very well covered, which could 
affect the quality of the applied Fourier series and consequently the mean magnitude. 
Supplementing the data for this star will possibly eliminate this problem in the future. 

Apart from AD~Cam, EW~Car is also rejected after $3\sigma$-clipping in all PW relations. 
There is no information about the reddening for that Cepheid in \citet{Fernie1995}, so it 
was not used for fitting the PL relation. Although the fit of the Fourier series for EW~Car  
(Figure~\ref{fig:fig1}) looks satisfactory, the quality of its light curve leaves much to 
be desired, as it was observed incidentally in the field containing the much brighter AQ~Car, 
meaning the exposure times were kept short to not saturate that target, resulting in lower-quality 
light curves for EW~Car. Ideally, a~bigger telescope should be employed to provide data of 
higher quality in the future. Nevertheless, the current data are good enough to obtain good 
values of the Wesenheit indices, which might suggest some reddening issue for EW~Car. 

In the case of $W^{gi}_{g}$ calculated based on the extinction coefficients from 
\citet{F99} besides AD~Cam and EW~Car, SV~Vul also has been rejected, for being too bright. 
This long-period Cepheid is another case of a~spectroscopic binary listed by \citet{Szabados1996}. 
They also list as spectroscopic binaries six additional stars common with our sample: 
AV~Sgr, LS~Pup, RT~Mus, RU~Sct, VW~Cen and YZ~Sgr. If we remove them from the sample then 
the slopes and intercepts of the PL and PW relations would still agree within the $1\sigma$ 
uncertainties. The identified binary Cepheids constitute about $7\%$ to $9\%$ of the sample 
of stars used for the PL and PW relations determination, respectively. 
As \citet{Karczmarek2023} show, the influence of binary Cepheids on the distance modulus 
(i.e. the difference between the zero-points of the PL relations in two galaxies) for this binarity 
fraction is very small for galaxies having similar percentages of binary Cepheids (see their 
Figure~7), and can be neglected. This means that there is no need to reject these stars from 
our sample. 

\subsection{Test for first overtone contamination} \label{ssec:firstovertone}

Although all Cepheid variables used by us in this analysis have been classified as 
fundamental mode pulsators \citep{PSU2021}, we performed a~test to confirm this. To 
avoid any significant contamination with first overtone Cepheids, we have adopted 
a~period cutoff of $5$~days, which corresponds to $\rm{log}P = 0.7$~days, and repeated 
the PL/PW relation fitting. The new slopes and zero-points agree within the $1\sigma$ 
uncertainties with the values from Tables~\ref{tab:plr} and \ref{tab:pwr} which assures 
us about the purity of our sample of stars regarding their pulsation modes. 

\subsection{The influence of the parallax ZP on the PL/PW relations} \label{ssec:ZPinfluence}

To evaluate the influence of the ZP offset calculated based on \citet{Lindegren2021a} 
on the found PL/PW relations we repeated the fitting without introducing parallax corrections 
as a~test. The slopes without a~ZP correction are steeper by slightly more than $1\sigma$, 
while the ZPs differ much more, and they are, on average, smaller by about $0.12$~mag 
for the PL relations and about $0.14$~mag in the case of the PW relations. 
The difference might be caused by the fact that the latter includes Cepheids from larger 
distances than the former. 

The same year when \citet[][hereafter L21]{Lindegren2021a} introduced their ZP offset 
corrections of Gaia parallaxes, also \citet[][hereafter G21]{Groenewegen2021} published his 
own, independent recipe. Both works use quasars and wide binaries as sources with known 
parallaxes for their calibrations. The main difference between the two approaches are the 
selection criteria: L21 use the $G$-magnitude, pseudocolor, and ecliptic latitude of a~given star, 
while G21 decided to use $G$-magnitude, $(G_{BP}-G_{RP})$ color, and for the spatial component 
HEALPix formalism was used to transform the sky coordinates into a~sky pixel (where the HEALPix 
level is a~parameter defining its resolution).  
We have calculated new parallax corrections according to the G21 instructions, where the 
mean new correction was about $-20\,\mu\rm{as}$, and have repeated the fitting of the PL/PW 
relations for our Cepheids.  
We obtained quite good agreement in the resulting coefficients with the ones calculated with 
the L21 corrections. The new slopes are slightly steeper, but they are well within the 
$1\sigma$ uncertainties, particularly for the absolute magnitudes and $W^{ri}_{r}$. For 
$W^{gr}_{r}$ and $W^{gi}_{g}$, however, the differences are higher (within about $2\sigma$). 
The intercepts for all relations are slightly smaller for G21, with the ZP shift 
being about $0.02$~mag, which agrees within $1\sigma$ with L21. 

\subsection{Dependence on metallicity?} \label{ssec:dep_met}

\begin{figure*} 
\plotone{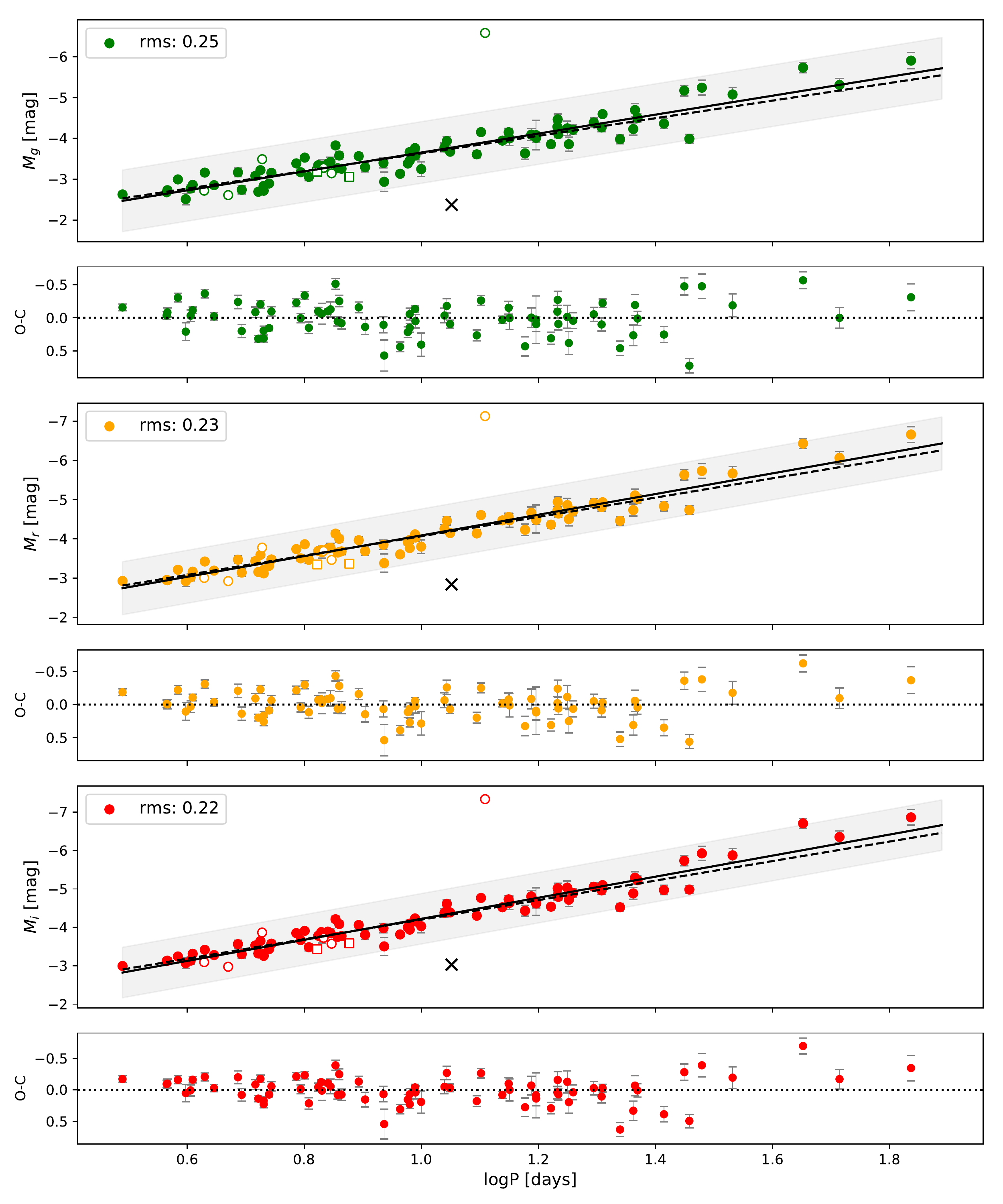}
\caption{The PL relations based on the mean reddenings from \citet{Fernie1995} 
and reddening vectors ($R_{\lambda}$) from \citet{Green2019}. 
Filled circles: Cepheids adopted for derivation of the PL relations; 
open circles: Cepheids with $RUWE>1.4$; 
open squares: Cepheids with poor-quality light curves;
black crosses: Cepheids rejected after $3\sigma$-clipping; 
black solid lines: the fit to Equation~(\ref{eq:plr}); 
black dashed line: the fit to Equation~(\ref{eq:abl_plr}); 
and the shaded areas: $\pm 3$rms.
\label{fig:plr}}
\end{figure*}

\begin{figure*} 
\plotone{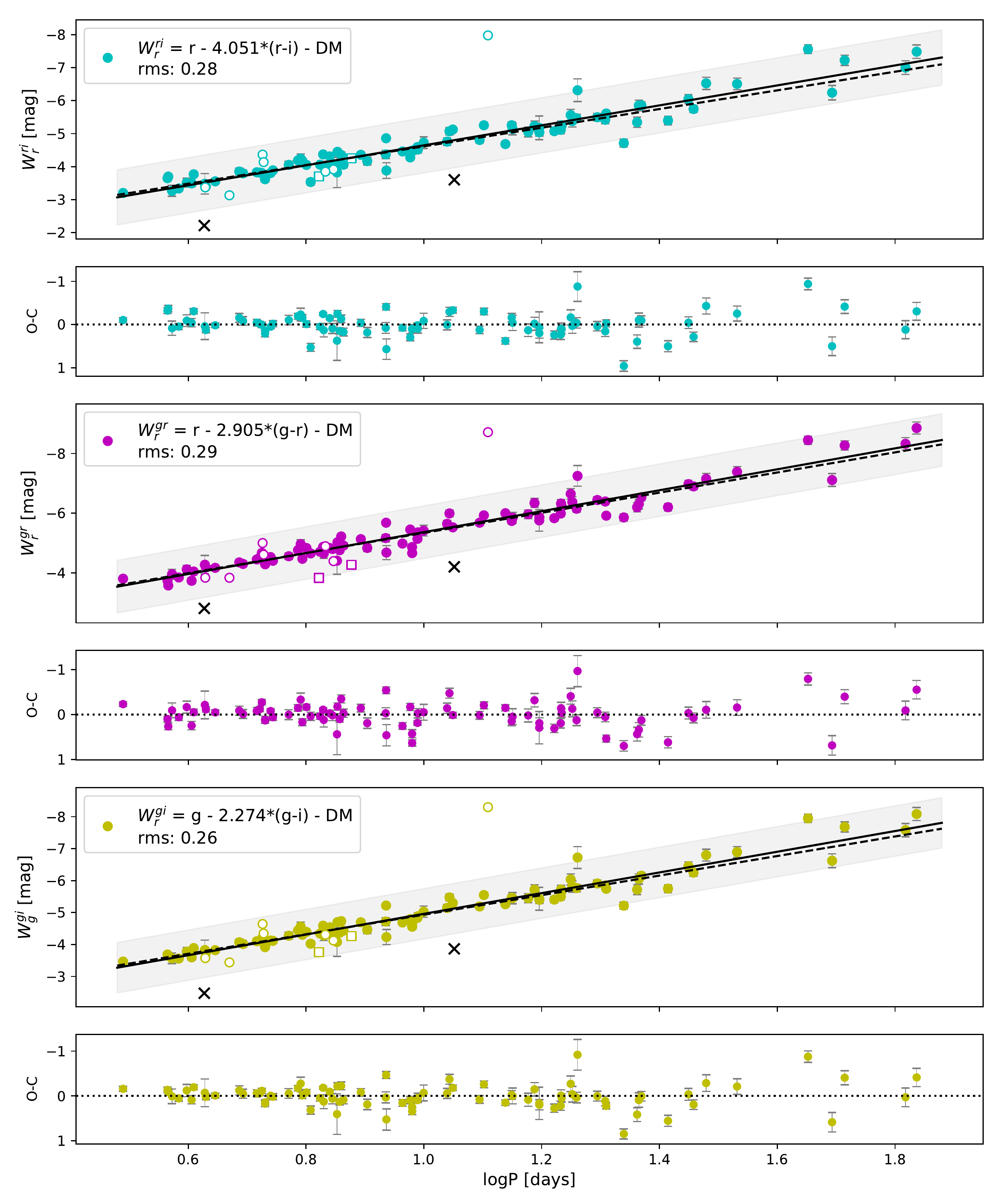}
\caption{The PW relation with Wesenheit magnitudes calculated based on the extinction 
coefficients from \citet{Green2019}. 
Filled circles: Cepheids used for derivation of the PW relations; 
open circles: Cepheids with $RUWE>1.4$; 
open squares: Cepheids with poor-quality light curves; 
black crosses: Cepheids rejected after $3\sigma$-clipping; 
black solid lines: the fit to Equation~(\ref{eq:pwr}); black dashed line: the fit to 
Equation~(\ref{eq:abl_pwr}); 
and the shaded areas: $\pm 3$rms.
\label{fig:pwr}}
\end{figure*}

\begin{figure*}
\gridline{\fig{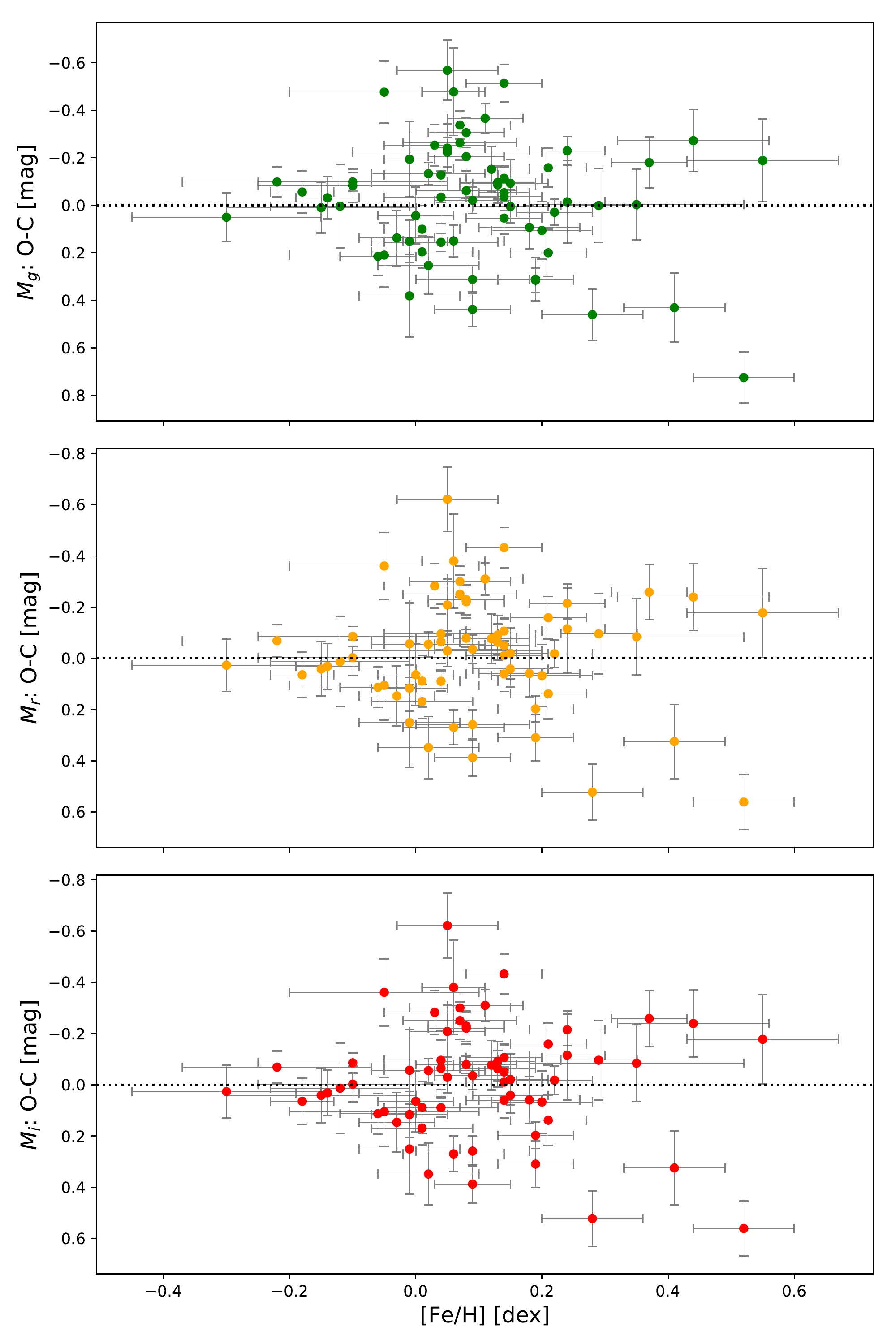}{0.5\textwidth}{(a)} 
	  \fig{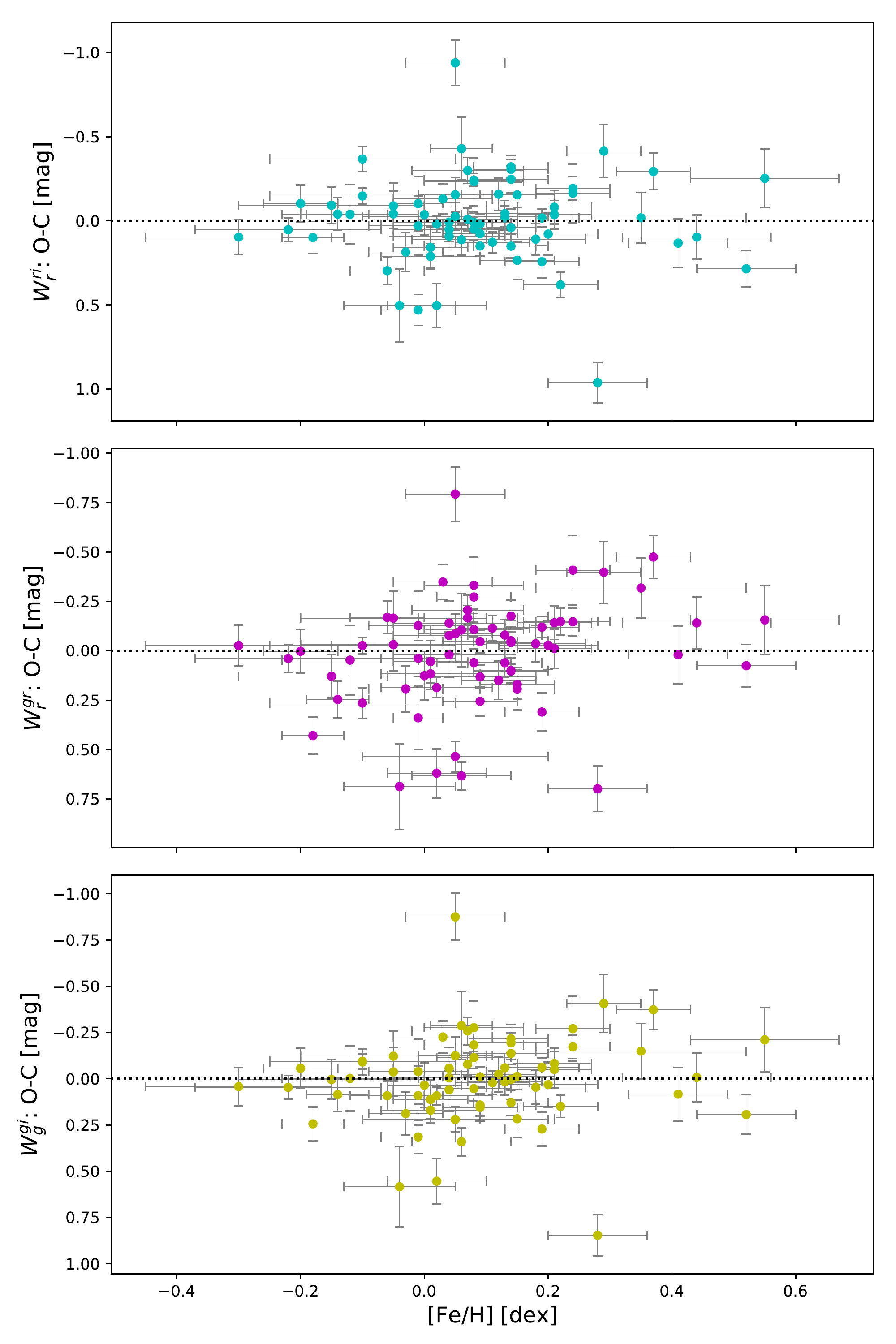}{0.5\textwidth}{(b)}
}
\caption{Residuals vs. metallicity for stars used in the fit of the PL (a) and PW (b) relations. 
\label{fig:dep_met}}
\end{figure*}

To check for any possible trend with metallicity, we adopted the metallicity values for the 
Cepheids in our sample from the summary provided by \citet{Breuval2021} in their 
Table~3. We then plotted the residuals of our PL and PW relations from Figures~\ref{fig:plr} 
and \ref{fig:pwr}, respectively, against metallicity values (see Figure~\ref{fig:dep_met}(a) 
and (b)). For the PL relations, metallicity information was available for $65$ stars, and for 
the PW relations for $68$ stars. For both cases, the metallicity range used is between $-0.30$ 
and $0.55$~dex, with an average and standard deviation of $0.09 \pm 0.16$~dex. 
The comparison of the PL/PW relations residuals with metallicity does not show any clear 
trends in any band, but it should be borne in mind that the range of metallicity used is 
rather narrow, and trends could possibly be revealed for a wider range of metallicities. 
Because of the small range of the metallicities of our Cepheids, we also decided not to 
investigate a~Period-Luminosity-Metallicity (PLZ) relation, however, such an analysis will 
be the subject of future projects. 

\subsection{Slope versus wavelength} \label{ssec:comp_with_lit}

The comparison of the slopes obtained by us using the two methods (linear and ABL) with other 
slope values presented in the literature, given in Figure~\ref{fig:slopes}, shows general 
agreement with both theoretical and observational results. The general trend is that the slopes 
become steeper for longer wavelengths. This is the behavior predicted by theoretical 
models \citep[e.g.,][]{Bono2010}, also observed in the data for any system. 
\citet{DiCriscienzo2013} provided a~theoretical scenario for the properties of classical Cepheids 
in SDSS filters. They used sets of nonlinear convective pulsation models 
computed for the typical chemical compositions of the Galaxy, transformed subsequently 
into the the SDSS photometric system. They noted the discrepancy of their resulting slopes 
with semiempirical and empirical ones from the literature, with the latter being systematically 
shallower. This is opposite in our case, as our results are systematically steeper than the 
theoretical ones. 
Simultaneously, the slopes obtained by us for the Pan-STARSS bands are shallower than those 
for the Johnson-Cousins bands \citep[see][]{Fouque2007,Breuval2021,Breuval2022}, which is, 
however, predicted by the aforementioned theoretical studies. 
The comparison of our slopes with the slopes for all fundamental mode classical Cepheids 
from M31 derived by \citet{Kodric2018} in the Pan-STARSS photometric system yields very 
satisfying agreement. Their results match our trends well, particularly the linear method, 
which was applied in that study. 

Worth noting is also the fact that the slopes obtained by us with the linear fit are steeper 
than those obtained with the ABL method, and the difference is more significant for longer 
wavelengths. The differences in the slope values (of either method) in the $r_{P1}$ and $i_{P1}$ 
filters calculated based on the reddening vectors from \citet{Green2019} and \citet{F99} are 
statistically insignificant, but they are largest for the $g_{P1}$ band. This behavior, however, 
is not surprising, as reddening laws differ the most from each other in the blue part of the 
spectrum. 


\subsection{Comparison of the PL relations with the literature} \label{ssec:comp_with_lit}

\begin{figure*} 
\plotone{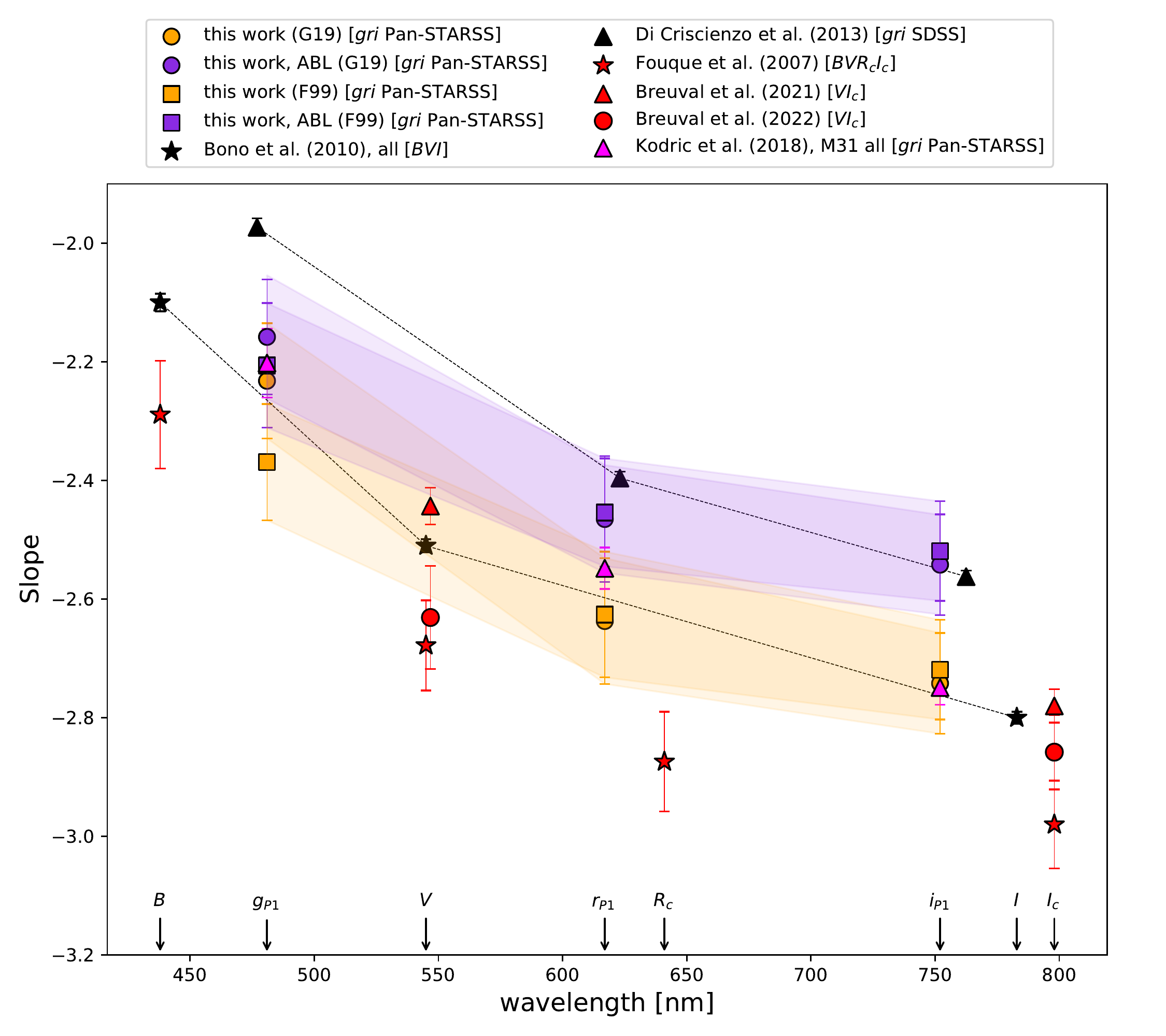}
\caption{Comparison of the PL relation slope at different wavelengths. 
Orange points mark the slopes obtained in this work from linear fitting, and the purple 
from the ABL method. Black points mark theoretical predictions for fundamental mode 
classical Cepheids, and the red points are obtained from fits to observations. 
Magenta triangles mark the slopes of the relations obtained for fundamental mode Cepheids 
from M31.
\label{fig:slopes}}
\end{figure*}

\begin{figure*} 
\gridline{\fig{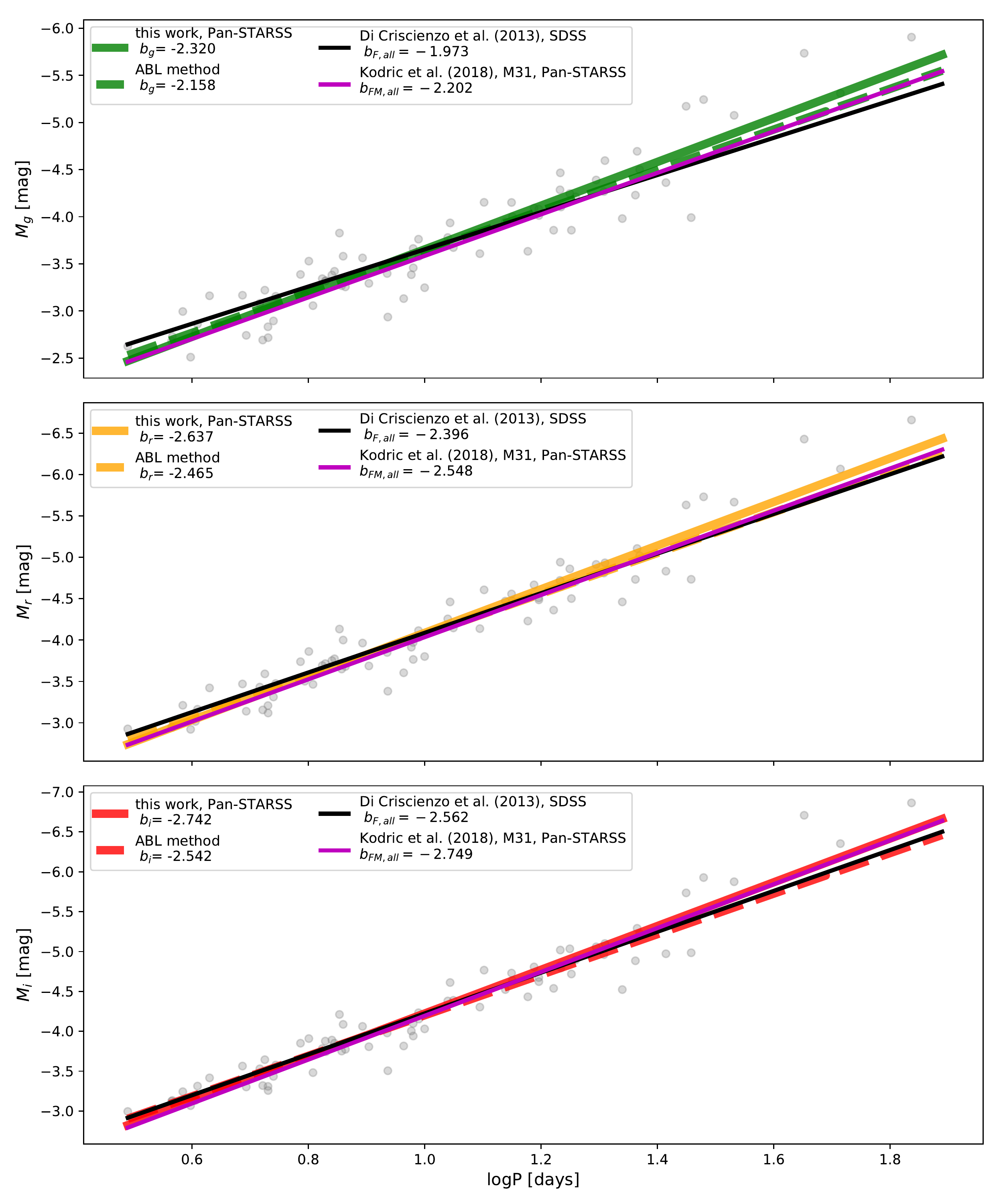}{0.5\textwidth}{(a)}
	  \fig{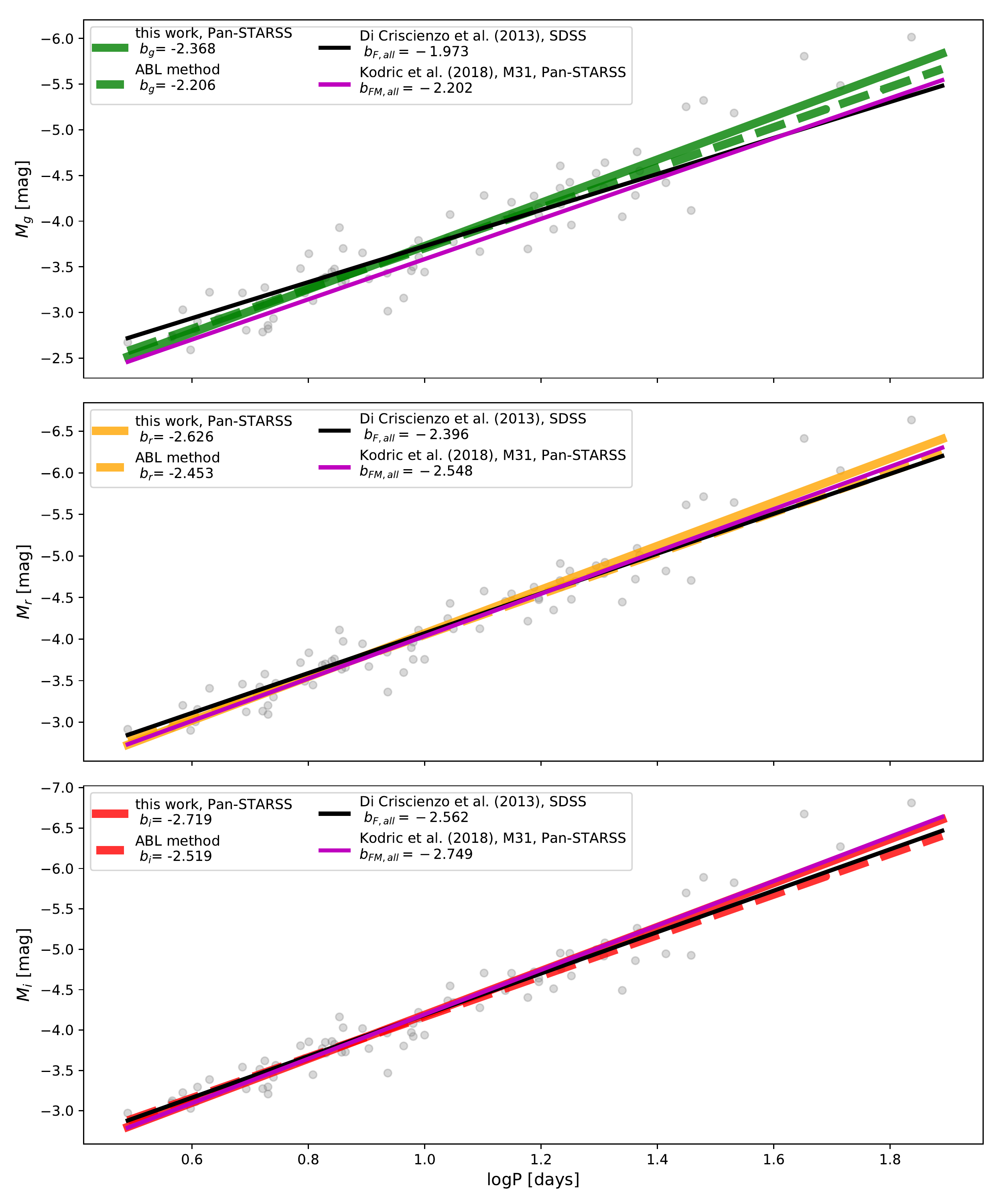}{0.5\textwidth}{(b)}
}
\gridline{\fig{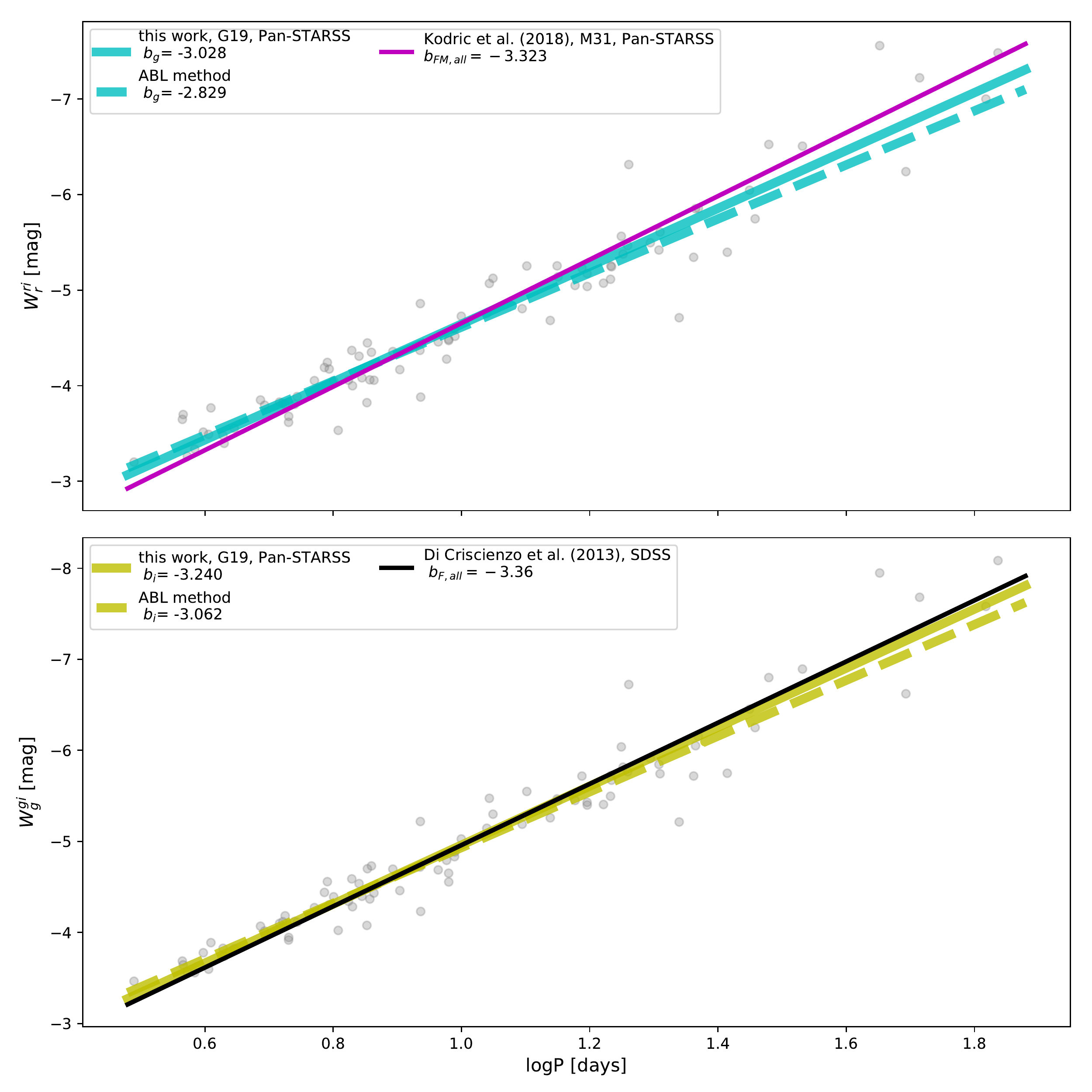}{0.5\textwidth}{(c)}
	  \fig{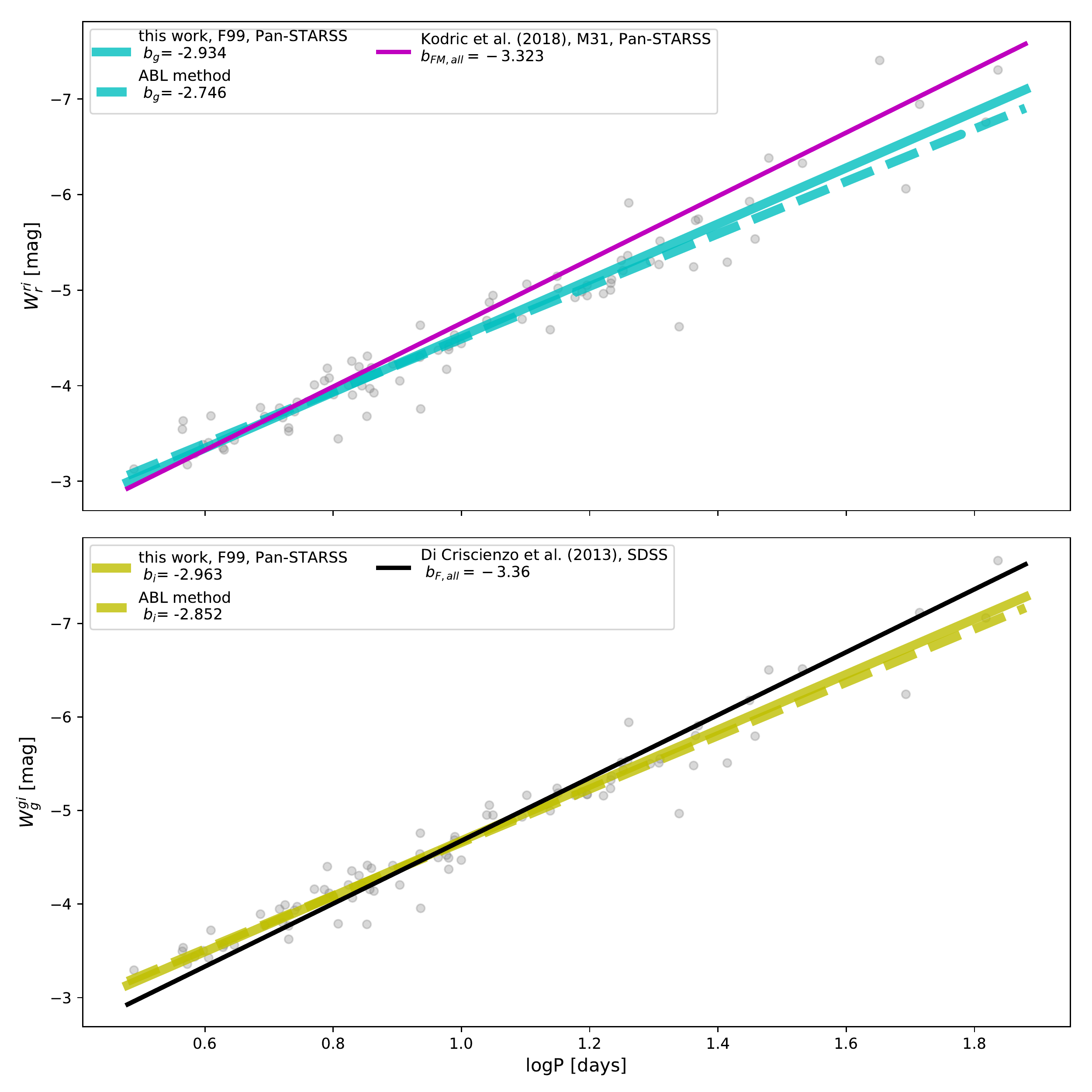}{0.5\textwidth}{(d)}
}
\caption{Comparison of the PL and PW relations obtained based on the reddening vectors 
from (a)-(c) \citet[][see Fig.~\ref{fig:plr}]{Green2019} and (b)-(d) \citet{F99} with the 
theoretical slopes from \citet[black, solid line]{DiCriscienzo2013} and the PL/PW relations 
from \citet[magenta, solid line]{Kodric2018} for M31 shifted by $\mu_{0\,M31} = 24.407$~mag 
from \citet{Li2021}.
\label{fig:plr_lit_comp}}
\end{figure*}

For the sake of comparison, we used the theoretical slopes from \citet[see their Table~2]{DiCriscienzo2013}  
predicted for all stars with $\log P \leq 2.0$ in the SDSS photometric system (marked also with 
black triangles in Figure~\ref{fig:slopes}), applied our zero-points, and plotted them with black, 
solid lines in Figure~\ref{fig:plr_lit_comp} over our PL relations from Table~\ref{tab:plr}. 
We also used the PL relations derived for M31 by \citet{Kodric2018} for fundamental mode Cepheids 
from the full range of periods given in their Table~3, shifted them by the distance modulus 
from \citet{Li2021} equal to $\mu_{0\,M31} = 24.407$~mag, and then overplotted with the magenta, 
solid line in Figure~\ref{fig:plr_lit_comp}(a) and (b). Analogously, we plotted the PW relation 
for M31 available for $W_{r}^{ri}$ and used the theoretical slope for $W_{g}^{gi}$ over the 
corresponding relations derived by us in Table~\ref{tab:pwr} (see Figure~\ref{fig:plr_lit_comp}(c) 
and (d)).

Such a~comparison shows very good agreement of our resulting PL relations with the theoretical 
predictions for the SDSS system (similar to Pan-STARSS), particularly in the $r_{P1}$ and $i_{P1}$ 
bands, while the $g_{P1}$ band shows the worst compliance, as our slopes in the case of the PL 
relations are noticeably steeper, while the $W_{g}^{gi}$ slopes are shallower than those reported 
by \citet{DiCriscienzo2013}. 
Very positive and encouraging is the comparison with the empirical relations for M31 obtained  
in the same Pan-STARSS photometric system, although the conclusion about the PL relations in the 
$g_{P1}$ band remains unchanged. 
This band shows the worst agreement, particularly for the reddening vectors obtained directly 
from the \citet{F99} reddening law. Also the $W_{r}^{ri}$ values calculated based on this law differ 
much from the $W_{r}^{ri}$ values from \citet{Kodric2018}, potentially resulting in smaller measured 
distances. 

As a~consequence, we recommend using the $g_{P1}$ band (and Wesenheit indices based on it) 
for distance scale purposes with caution. It is particularly sensitive to the applied reddening 
law, as well as characterized by the largest resulting uncertainties, among others, from the 
larger intrinsic spread of the instability strip in the blue bands. 
The $r_{P1}$ and $i_{P1}$ bands, however, are very promising tools for future applications,  
especially in the future LSST survey. 


\section{Summary} \label{sec:summary}

In this work we presented a~calibration of the PL relations in the Sloan -- Pans-STARRS 
$g_{P1}r_{P1}i_{P1}$ filters and $W_{r}^{ri}$, $W_{r}^{gr}$, and $W_{g}^{gi}$ Wesenheit 
indices for classical Cepheids from the MW pulsating in the fundamental mode. 
According to our knowledge, this was performed using MW Cepheids for the first time here. 
We obtained the data for our analysis using the 40-cm telescopes of the LCO Telescope 
Network in the three $g'r'i'$ Sloan filters, used parallaxes from the Gaia DR3 catalog, 
and ATLAS-REFCAT2 \citep{Tonry2018} to calibrate the photometry into the Pan-STARRS 
version of the Sloan photometric system. 
The obtained light curves, which we present in Figure~\ref{fig:fig1} in the Appendix~\ref{app:lc}, 
are generally well covered in pulsation phase, which allowed us to determine reliable mean 
magnitudes for a~sample of $96$ Cepheids used for this analysis. 
The homogeneous reddening values were adopted from \citet[][which, however, were not available 
for all Cepheids from our sample]{Fernie1995} and the extinction coefficients from 
\citet{Green2019} and calculated directly from the \citet{F99} reddening law.
After application of the selection criteria on the Gaia RUWE and GOF parameters and rejecting 
stars with poor-quality light curves, the number of stars used for the PL relations was 
$76$ (see Table~\ref{tab:plr}) and for the PW relation between $84$ and $85$ 
(see Table~\ref{tab:pwr}). The Wesenheit magnitudes were calculated based on the total to 
selective extinction ratios from two sources: \citet{Green2019} and \citet{F99}. 

The predominant contribution to the absolute magnitude uncertainties and subsequently 
the statistical errors of our PL/PW relations comes from the parallax uncertainties, while 
the mean magnitude errors are minor. The systematic uncertainties of the ZPs of our 
relations come from the Gaia parallax ZPs and were estimated to be $0.025$~mag, and the 
photometric ZP uncertainties, which we adopted as $0.01$~mag.

We have checked the behavior of the PL/PW relations in case of removing the confirmed 
binary Cepheids from our sample, and we did not notice any significant change. This test 
led us to the conclusion that there is no need for rejecting those stars from the fit. 
 
We performed a~comparison of the residuals of our PL/PW relations with the metallicities 
of those Cepheids, for which they were provided in Table~3 in \citet{Breuval2021}, and did 
not notice any trends. Because of the small range of available metallicities, we have decided 
not to perform a~fit to obtain the PLZ relation. To do such a~fit properly, 
a~much larger range of metallicities is required, preferably using Cepheids also from nearby 
galaxies, which will be the aim of projects in the future.

We compared our resulting slopes for the Galactic PL relations in the Pan-STARSS photometric 
system with theoretical slopes in the SDSS system \citep{DiCriscienzo2013}, as well as empirical 
ones in other passbands. We observe overall good agreement with the general trend that the slopes 
become steeper for longer wavelengths. Moreover, our PL relations, especially in the $r_{P1}$ 
and $i_{P1}$ bands, agree well with the empirical PL relations for M31 derived by \citet{Kodric2018}, 
which are also in the Pan-STARSS photometric system, while the agreement for the $g_{P1}$ band 
is the poorest. 

We believe the relations derived here will be a~useful tool for future distance determinations 
in the era of the upcoming LSST project, although we keep in mind that improvements are still 
easily possible. This could be done by extending the sample of used Cepheids, particularly 
by long-period Cepheids; by collecting data using homogeneous observations; by improving the 
coverage of the light curves; and by improving the extinction estimates of Cepheids.
Future Gaia data releases are also expected to provide parallaxes of even better quality.

\begin{longrotatetable}
\begin{deluxetable*}{hlclccchccc}
\tabletypesize{\scriptsize}
\tablenum{1}
\tablecaption{Sample of Milky Way Cepheids and their main parameters \label{tab:cepheids}}
\tablewidth{0pt}
\tablehead{
\nocolhead{Order} & \colhead{Star} & \colhead{Period} & \colhead{$\varpi_{DR3}$} & \colhead{RUWE} & 
\colhead{GOF} & \colhead{E(B-V)} & \nocolhead{Reference} & \colhead{$<g>$} & \colhead{$<r>$} & 
\colhead{$<i>$} \\
\nocolhead{number} &  & \colhead{(days)} & \colhead{(mas)} &  &  & 
\colhead{(mag)} &  & \colhead{(mag)} & \colhead{(mag)} & \colhead{(mag)} 
}
\decimalcolnumbers
\startdata
1 F    & AC~Mon &  8.0149308 & 0.3829 $\pm$ 0.0205 & 1.38 &  8.33 & 0.539 $\pm$ 0.035 & F95 & 10.574 $\pm$ 0.002 & 9.723 $\pm$ 0.002 & 9.277 $\pm$ 0.003 \\
2 F    & AD~Cam & 11.2630484 & 0.3150 $\pm$ 0.0192 & 1.19 &  5.34 & 0.929 $\pm$ 0.013 & F95 & 13.209 $\pm$ 0.048 & 11.954 $\pm$ 0.028 & 11.202 $\pm$ 0.020 \\
3 F    & AE~Vel &  7.1339448 & 0.3690 $\pm$ 0.0133 & 0.97 & -0.80 & 0.735 $\pm$ 0.058 & F95 & 10.769 $\pm$ 0.003 & 9.840 $\pm$ 0.004 & 9.316 $\pm$ 0.003 \\ 
4 F    & AG~Cru &  3.8372760 & 0.7581 $\pm$ 0.0219 & 1.02 &  0.53 & 0.257 $\pm$ 0.021 & F95 & 8.457 $\pm$ 0.008 & 8.022 $\pm$ 0.005 & 7.835 $\pm$ 0.007 \\ 
5 F    & AO~Aur &  6.7623732 & 0.2457 $\pm$ 0.0177 & 1.14 &  3.75 & 0.465 $\pm$ 0.050 & F95 & 11.264 $\pm$ 0.004 & 10.531 $\pm$ 0.003 & 10.165 $\pm$ 0.003 \\ 
6 F    & AQ~Car &  9.7695137 & 0.3611 $\pm$ 0.0171 & 1.07 &  1.77 & 0.179 $\pm$ 0.014 & F95 & 9.224 $\pm$ 0.003 & 8.618 $\pm$ 0.004 & 8.390 $\pm$ 0.004 \\ 
8 F    & AQ~Pup & 30.1496599 & 0.2939 $\pm$ 0.0248 & 1.18 &  5.07 & 0.565 $\pm$ 0.018 & F95 & 9.284 $\pm$ 0.003 & 8.317 $\pm$ 0.006 & 7.778 $\pm$ 0.008 \\ 
10 F   & AT~Pup &  6.6649973 & 0.6038 $\pm$ 0.0174 & 1.04 &  1.23 & 0.177 $\pm$ 0.012 & F95 & 8.338 $\pm$ 0.006 & 7.839 $\pm$ 0.006 & 7.641 $\pm$ 0.006 \\ 
11 F   & AV~Sgr & 15.4117020 & 0.4040 $\pm$ 0.0277 & 0.85 & -3.25 & 1.317 $\pm$ 0.029 & F95 & 12.231 $\pm$ 0.006 & 10.540 $\pm$ 0.004 & 9.601 $\pm$ 0.005 \\ 
13 F   & AY~Cen &  5.3094716 & 0.5741 $\pm$ 0.0157 & 0.95 & -1.45 & 0.380 $\pm$ 0.070 & F95 & 9.241 $\pm$ 0.003 & 8.548 $\pm$ 0.002 & 8.265 $\pm$ 0.002 \\ 
14 F   & BB~Sgr$^{b}$ &  6.6360805 & 1.1876 $\pm$ 0.0261 & 0.82 & -3.10 & 0.303 $\pm$ 0.012 & F95 & 7.448 $\pm$ 0.008 & 7.025 $\pm$ 0.009 & 6.753 $\pm$ 0.018 \\ 
15 F   & BF~Oph &  4.0675349 & 1.1892 $\pm$ 0.0264 & 0.84 & -4.98 & 0.278 $\pm$ 0.017 & F95 & 7.682 $\pm$ 0.014 & 7.144 $\pm$ 0.008 & 6.826 $\pm$ 0.006 \\ 
16 F   & BG~Vel &  6.9237971 & 1.0449 $\pm$ 0.0186 & 0.99 & -0.33 & 0.462 $\pm$ 0.012 & F95 & 8.051 $\pm$ 0.004 & 7.289 $\pm$ 0.004 & 6.871 $\pm$ 0.006 \\ 
17 F   & CD~Cyg & 17.0785209 & 0.3938 $\pm$ 0.0180 & 1.01 &  0.24 & 0.545 $\pm$ 0.022 & F95 & 9.540 $\pm$ 0.006 & 8.643 $\pm$ 0.010 & 8.215 $\pm$ 0.011 \\ 
19 F   & CE~Pup & 49.3224331 & 0.1138 $\pm$ 0.0113 & 0.82 & -5.45 & - & - & 12.475 $\pm$ 0.002 & 11.229 $\pm$ 0.002 & 10.550 $\pm$ 0.002 \\ 
21 F   & CN~Car &  4.9326581 & 0.3421 $\pm$ 0.0155 & 0.97 & -0.91 & 0.466 $\pm$ 0.052 & F95 & 11.129 $\pm$ 0.002 & 10.336 $\pm$ 0.002 & 9.891 $\pm$ 0.003 \\ 
126 F  & CN~Sct &  9.9939177 & 0.4022 $\pm$ 0.0326 & 0.94 & -1.25 & 1.407 $\pm$ 0.140 & F95 & 13.384 $\pm$ 0.004 & 11.638 $\pm$ 0.003 & 10.555 $\pm$ 0.003 \\ 
26 F   & CP~Cep & 17.8661365 & 0.2786 $\pm$ 0.0225 & 1.01 &  0.27 & 0.724 $\pm$ 0.048 & F95 & 11.312 $\pm$ 0.003 & 10.054 $\pm$ 0.003 & 9.398 $\pm$ 0.003 \\
27 F   & CV~Mon &  5.3786737 & 0.6010 $\pm$ 0.0163 & 1.10 &  2.17 & 0.750 $\pm$ 0.019 & F95 & 10.869 $\pm$ 0.002 & 9.832 $\pm$ 0.002 & 9.238 $\pm$ 0.003 \\
28 F   & CY~Car &  4.2659148 & 0.4274 $\pm$ 0.0123 & 0.93 & -2.17 & 0.435 $\pm$ 0.046 & F95 & 10.123 $\pm$ 0.002 & 9.494 $\pm$ 0.002 & 9.236 $\pm$ 0.002 \\ 
30 F   & DR~Vel & 11.1993194 & 0.5196 $\pm$ 0.0147 & 1.01 &  0.16 & 0.724 $\pm$ 0.026 & F95 & 10.143 $\pm$ 0.003 & 9.056 $\pm$ 0.004 & 8.375 $\pm$ 0.004 \\
31 F   & DX~Pyx &  3.7372519 & 0.2246 $\pm$ 0.0170 & 1.13 &  3.71 & - & - & 12.084 $\pm$ 0.002 & 11.368 $\pm$ 0.002 & 11.025 $\pm$ 0.002 \\
32 F   & EK~Mon &  3.9579526 & 0.3762 $\pm$ 0.0233 & 1.16 &  2.98 & 0.582 $\pm$ 0.003 & F95 & 11.538 $\pm$ 0.002 & 10.634 $\pm$ 0.002 & 10.134 $\pm$ 0.002 \\
33 F   & ER~Aur & 15.7002656 & 0.1287 $\pm$ 0.0213 & 1.16 &  4.45 & 0.491 $\pm$ 0.028 & F95 & 11.995 $\pm$ 0.002 & 11.150 $\pm$ 0.002 & 10.688 $\pm$ 0.002 \\
7 F    & EW~Car &  4.2390336 & 0.1745 $\pm$ 0.0164 & 0.99 & -0.36 & - & - & 14.440 $\pm$ 0.031 & 13.648 $\pm$ 0.014 & 13.227 $\pm$ 0.016 \\ 
35 F   & FM~Aql &  6.1142663 & 0.0145 $\pm$ 0.0283 & 1.26 &  5.44 & 0.676 $\pm$ 0.020 & F95 & 8.817 $\pm$ 0.004 & 7.893 $\pm$ 0.009 & 7.371 $\pm$ 0.006 \\
36 F   & FN~Aql &  9.4815651 & 0.7362 $\pm$ 0.0270 & 1.13 &  2.88 & 0.517 $\pm$ 0.009 & F95 & 8.990 $\pm$ 0.004 & 8.023 $\pm$ 0.004 & 7.619 $\pm$ 0.004 \\
37 F   & FN~Vel$^{a}$ &  5.3241672 & 0.3512 $\pm$ 0.0236 & 1.700 & 15.98 & - & - & 10.812 $\pm$ 0.002 & 9.906 $\pm$ 0.003 & 9.413 $\pm$ 0.002 \\
40 F   & GH~Cyg &  7.8163294 & 0.4171 $\pm$ 0.0158 & 1.07 &  2.25 & 0.647 $\pm$ 0.024 & F95 & 10.475 $\pm$ 0.004 & 9.526 $\pm$ 0.004 & 9.036 $\pm$ 0.003 \\
42 F   & GQ~Ori &  8.6165158 & 0.4082 $\pm$ 0.0229 & 0.87 & -2.97 & 0.238 $\pm$ 0.014 & F95 & 9.335 $\pm$ 0.004 & 8.681 $\pm$ 0.002 & 8.408 $\pm$ 0.006 \\
43 F   & GX~Car &  7.1968800 & 0.4590 $\pm$ 0.0145 & 1.02 &  0.48 & 0.404 $\pm$ 0.009 & F95 & 9.759 $\pm$ 0.003 & 9.035 $\pm$ 0.002 & 8.688 $\pm$ 0.004 \\
44 F   & GY~Sge & 51.8140890 & 0.3421 $\pm$ 0.0246 & 0.95 & -1.61 & 1.258 $\pm$ 0.050 & F95 & 11.179 $\pm$ 0.002 & 9.356 $\pm$ 0.002 & 8.307 $\pm$ 0.003 \\
46 F   & HQ~Per &  8.6387834 & 0.2341 $\pm$ 0.0254 & 1.30 &  7.74 & 0.571 $\pm$ 0.021 & F95 & 12.106 $\pm$ 0.002 & 11.176 $\pm$ 0.002 & 10.706 $\pm$ 0.002 \\
47 F   & HW~Car &  9.1992766 & 0.3972 $\pm$ 0.0134 & 0.94 & -1.52 & 0.193 $\pm$ 0.050 & F95 & 9.512 $\pm$ 0.001 & 8.875 $\pm$ 0.002 & 8.547 $\pm$ 0.002 \\
121 F  & IP~Car &  7.1213761 & 0.0801 $\pm$ 0.0165 & 1.19 &  4.78 & - & - & 14.805 $\pm$ 0.018 & 13.850 $\pm$ 0.016 & 13.309 $\pm$ 0.020 \\
48 F   & IT~Car$^{b}$ &  7.5330469 & 0.7017 $\pm$ 0.0223 & 1.08 &  2.23 & 0.226 $\pm$ 0.017 & F95 & 8.460 $\pm$ 0.004 & 7.958 $\pm$ 0.006 & 7.602 $\pm$ 0.006 \\
51 F   & KN~Cen & 34.0213123 & 0.2506 $\pm$ 0.0201 & 1.03 &  1.01 & 0.775 $\pm$ 0.043 & F95 & 10.492 $\pm$ 0.002 & 9.244 $\pm$ 0.002 & 8.566 $\pm$ 0.002 \\
52 F   & KQ~Sco & 28.7030114 & 0.4717 $\pm$ 0.0232 & 0.91 & -2.55 & 0.906 $\pm$ 0.044 & F95 & 10.637 $\pm$ 0.004 & 9.126 $\pm$ 0.003 & 8.326 $\pm$ 0.003 \\
9 F    & LS~Pup & 14.1472871 & 0.2139 $\pm$ 0.0173 & 1.25 &  7.25 & 0.481 $\pm$ 0.010 & F95 & 10.940 $\pm$ 0.002 & 10.060 $\pm$ 0.002 & 9.603 $\pm$ 0.002 \\ 
20 F   & OGLE~GD-CEP-189 &  4.2483435 & 0.0922 $\pm$ 0.0130 & 1.08 &  2.19 & - & - & 14.677 $\pm$ 0.006 & 13.711 $\pm$ 0.004 & 13.214 $\pm$ 0.004 \\ 
54 F   & QY~Cen & 17.7542664 & 0.2931 $\pm$ 0.0235 & 1.02 &  0.82 & 1.290 $\pm$ 0.230 & F95 & 12.685 $\pm$ 0.005 & 10.978 $\pm$ 0.003 & 10.021 $\pm$ 0.003 \\
55 F   & RT~Mus &  3.0861175 & 0.6786 $\pm$ 0.0160 & 0.94 & -1.60 & 0.343 $\pm$ 0.037 & F95 & 9.351 $\pm$ 0.004 & 8.759 $\pm$ 0.004 & 8.483 $\pm$ 0.004 \\
56 F   & RU~Sct & 19.7042782 & 0.5256 $\pm$ 0.0261 & 0.87 & -2.60 & 0.972 $\pm$ 0.018 & F95 & 10.221 $\pm$ 0.004 & 8.873 $\pm$ 0.005 & 8.139 $\pm$ 0.005 \\
57 F   & RY~Sco & 20.3223325 & 0.7639 $\pm$ 0.0353 & 0.73 & -6.59 & 0.696 $\pm$ 0.047 & F95 & 8.619 $\pm$ 0.005 & 7.486 $\pm$ 0.009 & 6.913 $\pm$ 0.015 \\
58 F   & RY~Vel & 28.1341673 & 0.3759 $\pm$ 0.0227 & 1.08 &  1.97 & 0.573 $\pm$ 0.013 & F95 & 8.847 $\pm$ 0.003 & 7.901 $\pm$ 0.005 & 7.451 $\pm$ 0.006 \\
60 F   & RZ~CMa$^{a}$ &  4.2550780 & 0.5860 $\pm$ 0.0260 & 1.63 & 14.69 & 0.505 $\pm$ 0.021 & F95 & 10.113 $\pm$ 0.002 & 9.399 $\pm$ 0.002 & 9.002 $\pm$ 0.002 \\
62 F   & RZ~Vel & 20.4009397 & 0.6615 $\pm$ 0.0184 & 1.24 &  6.05 & 0.320 $\pm$ 0.012 & F95 & 7.360 $\pm$ 0.012 & 6.751 $\pm$ 0.009 & 6.390 $\pm$ 0.012 \\
63 F   & S~Nor &  9.7544164 & 1.0989 $\pm$ 0.0241 & 0.88 & -3.74 & 0.194 $\pm$ 0.008 & F95 & 6.675 $\pm$ 0.005 & 6.158 $\pm$ 0.004 & 5.922 $\pm$ 0.003 \\
64 F   & SS~Sct &  3.6713516 & 0.9342 $\pm$ 0.0250 & 0.84 & -3.31 & 0.362 $\pm$ 0.023 & F95 & 8.665 $\pm$ 0.004 & 8.088 $\pm$ 0.004 & 7.696 $\pm$ 0.008 \\
65 F   & ST~Tau &  4.0343040 & 0.9162 $\pm$ 0.0374 & 1.35 & 10.54 & 0.349 $\pm$ 0.006 & F95 & 8.574 $\pm$ 0.006 & 8.032 $\pm$ 0.007 & 7.703 $\pm$ 0.009 \\
66 F   & SU~Cru$^{a}$ & 12.8491103 & 0.1783 $\pm$ 0.1597 & 1.52 & 10.50 & 1.027 $\pm$ 0.056 & F95 & 10.556 $\pm$ 0.003 & 9.141 $\pm$ 0.003 & 8.308 $\pm$ 0.003 \\
69 F   & S~Vul & 68.6510481 & 0.2373 $\pm$ 0.0222 & 1.04 &  1.13 & 0.782 $\pm$ 0.051 & F95 & 9.804 $\pm$ 0.002 & 8.387 $\pm$ 0.005 & 7.709 $\pm$ 0.006 \\
70 F   & SV~Vel & 14.0980332 & 0.4342 $\pm$ 0.0193 & 1.02 &  0.57 & 0.400 $\pm$ 0.025 & F95 & 8.983 $\pm$ 0.003 & 8.238 $\pm$ 0.003 & 7.823 $\pm$ 0.003 \\
71 F   & SV~Vul & 44.8942247 & 0.4017 $\pm$ 0.0233 & 1.20 &  6.14 & 0.504 $\pm$ 0.026 & F95 & 7.912 $\pm$ 0.011 & 6.791 $\pm$ 0.012 & 6.206 $\pm$ 0.006 \\
73 F   & SW~Vel & 23.4069247 & 0.4133 $\pm$ 0.0202 & 1.05 &  1.34 & 0.360 $\pm$ 0.010 & F95 & 8.609 $\pm$ 0.005 & 7.784 $\pm$ 0.007 & 7.359 $\pm$ 0.004 \\
74 F   & SX~Car &  4.8600155 & 0.5149 $\pm$ 0.0240 & 1.25 &  6.19 & 0.344 $\pm$ 0.028 & F95 & 9.412 $\pm$ 0.002 & 8.818 $\pm$ 0.003 & 8.515 $\pm$ 0.003 \\
75 F   & SX~Vel &  9.5506149 & 0.5006 $\pm$ 0.0206 & 1.02 &  0.60 &  0.252 $\pm$ 0.015 & F95 & 8.673 $\pm$ 0.004 & 8.152 $\pm$ 0.006 & 7.872 $\pm$ 0.008 \\
76 F   & SZ~Aql & 17.1413213 & 0.5255 $\pm$ 0.0220 & 0.94 & -1.21 & 0.588 $\pm$ 0.023 & F95 & 9.238 $\pm$ 0.003 & 8.198 $\pm$ 0.004 & 7.693 $\pm$ 0.004 \\
77 F   & T~Ant &  5.8983587 & 0.3124 $\pm$ 0.0156 & 1.18 &  6.30 & - & - & 9.572 $\pm$ 0.003 & 9.162 $\pm$ 0.003 & 8.992 $\pm$ 0.003 \\
78 F   & TT~Aql & 13.7547268 & 0.9975 $\pm$ 0.0246 & 1.08 &  1.42 & 0.518 $\pm$ 0.025 & F95 & 7.773 $\pm$ 0.010 & 6.809 $\pm$ 0.008 & 6.442 $\pm$ 0.011 \\
79 F   & TW~CMa &  6.9955097 & 0.3839 $\pm$ 0.0204 & 1.15 &  3.46 & 0.398 $\pm$ 0.035 & F95 & 9.974 $\pm$ 0.003 & 9.284 $\pm$ 0.003 & 8.966 $\pm$ 0.003 \\
80 F   & TX~Cen & 17.0979569 & 0.3324 $\pm$ 0.0200 & 0.94 & -1.89 & 1.001 $\pm$ 0.040 & F95 & 11.236 $\pm$ 0.003 & 9.913 $\pm$ 0.003 & 9.228 $\pm$ 0.003 \\
127 F  & TY~Sct & 11.0540745 & 0.3586 $\pm$ 0.0179 & 0.91 & -2.00 & 0.989 $\pm$ 0.018 & F95 & 11.563 $\pm$ 0.002 & 10.199 $\pm$ 0.002 & 9.448 $\pm$ 0.002 \\
81 F   & U~Aql$^{a}$ &  7.0249035 & 1.7648 $\pm$ 0.0962 & 3.09 & 27.78 & 0.401 $\pm$ 0.010 & F95 & 6.949 $\pm$ 0.012 & 6.290 $\pm$ 0.017 & 5.935 $\pm$ 0.017 \\
82 F   & U~Nor & 12.6452592 & 0.6251 $\pm$ 0.0213 & 0.98 & -0.40 & 0.923 $\pm$ 0.040 & F95 & 9.920 $\pm$ 0.003 & 8.684 $\pm$ 0.003 & 7.964 $\pm$ 0.005 \\
83 F   & U~Sgr &  6.745226 & 1.6049 $\pm$ 0.0247 & 0.85 & -2.87 & 0.434 $\pm$ 0.007 & F95 & 7.090 $\pm$ 0.002 & 6.325 $\pm$ 0.004 & 5.900 $\pm$ 0.003 \\
84 F   & UW~Car$^{a}$ &  5.3458320 & 0.4471 $\pm$ 0.0236 & 1.62 & 15.04 & 0.455 $\pm$ 0.008 & F95 & 9.762 $\pm$ 0.003 & 9.090 $\pm$ 0.002 & 8.725 $\pm$ 0.002 \\
88 F   & UX~Car &  3.6821720 & 0.6532 $\pm$ 0.0210 & 1.02 &  0.47 & 0.109 $\pm$ 0.024 & F95 & 8.556 $\pm$ 0.004 & 8.247 $\pm$ 0.008 & 7.995 $\pm$ 0.003 \\
89 F   & UY~Car &  5.5437621 & 0.4548 $\pm$ 0.0148 & 0.94 & -1.63 & 0.200 $\pm$ 0.024 & F95 & 9.217 $\pm$ 0.002 & 8.728 $\pm$ 0.003 & 8.505 $\pm$ 0.002 \\
93 F   & UZ~Car &  5.2046310 & 0.4009 $\pm$ 0.0141 & 0.95 & -1.47 & 0.227 $\pm$ 0.036 & F95 & 9.654 $\pm$ 0.003 & 9.112 $\pm$ 0.004 & 8.876 $\pm$ 0.003 \\
94 F   & V1162~Aql &  5.3757961 & 0.8231 $\pm$ 0.0252 & 0.95 & -1.24 & 0.196 $\pm$ 0.012 & F95 & 8.239 $\pm$ 0.010 & 7.697 $\pm$ 0.009 & 7.477 $\pm$ 0.007 \\
95 F   & V1496~Aql & 65.7307917 & 0.2829 $\pm$ 0.0270 & 1.01 &  0.33 & - & - & 11.185 $\pm$ 0.003 & 9.452 $\pm$ 0.004 & 8.536 $\pm$ 0.004 \\
98 F   & V1803~Aql &  8.6279483 & 0.6085 $\pm$ 0.0210 & 0.94 & -1.62 & - & - & 11.179 $\pm$ 0.002 & 9.699 $\pm$ 0.002 & 8.840 $\pm$ 0.002 \\
99 F   & V1804~Aql & 18.2360555 & 0.2532 $\pm$ 0.0398 & 0.97 & -0.86 & - & - & 15.271 $\pm$ 0.017 & 12.829 $\pm$ 0.009 & 11.308 $\pm$ 0.010 \\
100 F  & V336~Aql &  7.3041815 & 0.5086 $\pm$ 0.0210 & 1.10 &  1.86 & 0.670 $\pm$ 0.012 & F95 & 10.428 $\pm$ 0.003 & 9.436 $\pm$ 0.004 & 8.936 $\pm$ 0.005 \\
101 F  & V496~Aql$^{a}$ &  6.8070785 & 0.9769 $\pm$ 0.0377 & 1.56 & 10.90 & 0.453 $\pm$ 0.020 & F95 & 8.272 $\pm$ 0.003 & 7.475 $\pm$ 0.003 & 7.161 $\pm$ 0.005 \\
102 F  & V496~Cen &  4.4241128 & 0.5629 $\pm$ 0.0143 & 0.94 & -1.77 & 0.616 $\pm$ 0.033 & F95 & 10.432 $\pm$ 0.003 & 9.574 $\pm$ 0.003 & 9.110 $\pm$ 0.002 \\
103 F  & V600~Aql &  7.2388599 & 0.5234 $\pm$ 0.0208 & 1.13 &  2.54 & 0.864 $\pm$ 0.007 & F95 & 10.682 $\pm$ 0.003 & 9.531 $\pm$ 0.004 & 8.920 $\pm$ 0.005 \\
104 F  & V733~Aql &  6.1787826 & 0.2441 $\pm$ 0.0160 & 0.99 & -0.39 & - & - & 10.323 $\pm$ 0.004 & 9.754 $\pm$ 0.004 & 9.523 $\pm$ 0.003 \\
105 F  & V~Cen &  5.4940002 & 1.4089 $\pm$ 0.0243 & 1.06 &  1.80 & 0.282 $\pm$ 0.017 & F95 & 7.294 $\pm$ 0.009 & 6.637 $\pm$ 0.006 & 6.344 $\pm$ 0.004 \\
106 F  & VW~Cen & 15.0368116 & 0.2598 $\pm$ 0.0173 & 1.06 &  1.96 & 0.451 $\pm$ 0.023 & F95 & 10.785 $\pm$ 0.002 & 9.806 $\pm$ 0.003 & 9.330 $\pm$ 0.003 \\
107 F  & VW~Cru &  5.2651814 & 0.7381 $\pm$ 0.0175 & 0.85 & -4.05 & 0.681 $\pm$ 0.049 & F95 & 10.220 $\pm$ 0.003 & 9.179 $\pm$ 0.003 & 8.602 $\pm$ 0.004 \\
110 F  & VZ~Pup & 23.1752288 & 0.2198 $\pm$ 0.0162 & 1.24 &  7.15 & 0.461 $\pm$ 0.019 & F95 & 10.120 $\pm$ 0.005 & 9.317 $\pm$ 0.004 & 8.853 $\pm$ 0.005 \\
111 F  & WW~Car$^{a}$ &  4.6768391 & 0.5269 $\pm$ 0.0200 & 1.48 & 12.53 & 0.408 $\pm$ 0.009 & F95 & 10.131 $\pm$ 0.003 & 9.472 $\pm$ 0.002 & 9.173 $\pm$ 0.002 \\
112 F  & WZ~Car & 23.0180376 & 0.2843 $\pm$ 0.0202 & 1.38 & 10.09 & 0.379 $\pm$ 0.007 & F95 & 9.755 $\pm$ 0.004 & 8.930 $\pm$ 0.005 & 8.549 $\pm$ 0.004 \\
113 F  & WZ~Sgr & 21.8532997 & 0.6119 $\pm$ 0.0305 & 0.94 & -1.45 & 0.486 $\pm$ 0.027 & F95 & 8.693 $\pm$ 0.004 & 7.801 $\pm$ 0.010 & 7.444 $\pm$ 0.011 \\
114 F  & X~Cru &  6.2201526 & 0.6541 $\pm$ 0.0209 & 0.95 & -1.67 & 0.313 $\pm$ 0.020 & F95 & 8.787 $\pm$ 0.006 & 8.189 $\pm$ 0.006 & 7.833 $\pm$ 0.006 \\
115 F  & X~Pup & 25.9722922 & 0.3970 $\pm$ 0.0221 & 1.04 &  1.19 & 0.421 $\pm$ 0.016 & F95 & 9.036 $\pm$ 0.005 & 8.209 $\pm$ 0.007 & 7.814 $\pm$ 0.009 \\
116 F  & X~Vul &  6.3196496 & 0.8642 $\pm$ 0.0237 & 1.06 &  2.18 & 0.824 $\pm$ 0.022 & F95 & 9.513 $\pm$ 0.002 & 8.482 $\pm$ 0.005 & 7.935 $\pm$ 0.007 \\
117 F  & XX~Car & 15.7070571 & 0.3049 $\pm$ 0.0159 & 1.07 &  1.95 & 0.373 $\pm$ 0.011 & F95 & 9.800 $\pm$ 0.007 & 9.011 $\pm$ 0.004 & 8.648 $\pm$ 0.004 \\
118 F  & XX~Cen & 10.9526746 & 0.5701 $\pm$ 0.0288 & 1.24 &  9.26 & 0.261 $\pm$ 0.013 & F95 & 8.304 $\pm$ 0.005 & 7.606 $\pm$ 0.006 & 7.324 $\pm$ 0.008 \\
119 F  & XX~Sgr &  6.4243335 & 0.7239 $\pm$ 0.0298 & 1.10 &  2.13 & 0.524 $\pm$ 0.017 & F95 & 9.378 $\pm$ 0.005 & 8.526 $\pm$ 0.002 & 8.191 $\pm$ 0.005 \\
120 F  & XY~Car & 12.4360971 & 0.3904 $\pm$ 0.0150 & 1.07 &  2.08 & 0.420 $\pm$ 0.013 & F95 & 9.823 $\pm$ 0.005 & 8.937 $\pm$ 0.006 & 8.517 $\pm$ 0.005 \\
122 F  & XZ~Car & 16.6528143 & 0.4729 $\pm$ 0.0198 & 1.05 &  1.38 & 0.396 $\pm$ 0.028 & F95 & 9.079 $\pm$ 0.005 & 8.238 $\pm$ 0.006 & 7.822 $\pm$ 0.005 \\
123 F  & YZ~Car & 18.1677926 & 0.3575 $\pm$ 0.0197 & 1.17 &  4.71 & 0.345 $\pm$ 0.042 & F95 & 9.163 $\pm$ 0.003 & 8.375 $\pm$ 0.005 & 7.978 $\pm$ 0.005 \\
125 F  & YZ~Sgr &  9.5537875 & 0.8596 $\pm$ 0.0267 & 0.96 & -0.75 & 0.307 $\pm$ 0.007 & F95 & 7.886 $\pm$ 0.004 & 7.318 $\pm$ 0.005 & 6.957 $\pm$ 0.015 \\
\enddata
\tablecomments{Star: name of the Cepheid; Period: period of the Cepheid adopted from \citet{PSU2021}; 
$\varpi_{DR3}$: parallax from the Gaia~DR3 catalog corrected with \citet{Lindegren2021a} corrections; 
RUWE: renormalized unit weight error from the Gaia~DR3 catalog; 
GOF: goodness-of-fit from the Gaia~DR3 catalog; 
E(B-V): reddening value from \citet{Fernie1995}; 
and $<g>$, $<r>$, $<i>$: mean magnitude from Fourier series fitting for the Pan-STARRS $g_{P1}r_{P1}i_{P1}$ 
filters, respectively. \\
$^{a}$ Star rejected based on the RUWE and GOF parallax quality parameters given by the Gaia~DR3. \\ 
$^{b}$ Star rejected because of the poor quality of its light curves. 
}
\end{deluxetable*}
\end{longrotatetable}

\begin{deluxetable*}{llcccc}
\tabletypesize{\scriptsize}
\tablewidth{0pt} 
\tablenum{2}
\tablecaption{The determined Period-Luminosity relations for Galactic classical Cepheids \label{tab:plr}}
\tablehead{
\colhead{Source} & \colhead{Band} & \colhead{$a_{\lambda}$} & \colhead{$b_{\lambda}$} 
& \colhead{rms} & \colhead{N}  
}
\colnumbers
\startdata 
\multicolumn{5}{l}{Period-Luminosity relation: $M_{\lambda} = a_{\lambda}(\rm{logP - 1}) + b_{\lambda}$} \\ 
\hline
{                 } & {$g$} & -2.320 $\pm$ 0.097 & -3.654 $\pm$ 0.029 & 0.25 & 76 \\ 
{\citet{Green2019}} & {$r$} & -2.637 $\pm$ 0.106 & -4.088 $\pm$ 0.032 & 0.23 & 76 \\
{                 } & {$i$} & -2.742 $\pm$ 0.085 & -4.223 $\pm$ 0.025 & 0.22 & 76 \\ 
\hline 
{           } & {$g$} & -2.369 $\pm$ 0.098 & -3.727 $\pm$ 0.029 & 0.25 & 76 \\ 
{\citet{F99}} & {$r$} & -2.626 $\pm$ 0.106 & -4.071 $\pm$ 0.032 & 0.22 & 76 \\
{           } & {$i$} & -2.719 $\pm$ 0.084 & -4.189 $\pm$ 0.025 & 0.22 & 76 \\ 
\hline
\multicolumn{5}{l}{Astrometry-Based Luminosity: $ABL_{\lambda} = 10^{0.2[a_{\lambda}(logP - 1) + b_{\lambda}]}$} \\ 
\hline
{                 } & {$g$} & -2.158 $\pm$ 0.104 & -3.631 $\pm$ 0.030 & 0.26 & 76 \\ 
{\citet{Green2019}} & {$r$} & -2.465 $\pm$ 0.091 & -4.067 $\pm$ 0.026 & 0.23 & 76 \\ 
{                 } & {$i$} & -2.542 $\pm$ 0.084 & -4.201 $\pm$ 0.024 & 0.23 & 76 \\  
\hline
{           } & {$g$} & -2.206 $\pm$ 0.105 & -3.705 $\pm$ 0.030 & 0.26 & 76 \\ 
{\citet{F99}} & {$r$} & -2.454 $\pm$ 0.091 & -4.050 $\pm$ 0.026 & 0.23 & 76 \\ 
{           } & {$i$} & -2.519 $\pm$ 0.084 & -4.166 $\pm$ 0.024 & 0.23 & 76 \\  
\enddata
\tablecomments{ 
Source: source of extinction vector ($R_{\lambda}$); 
Band: the Pan-STARRS $g_{P1}r_{P1}i_{P1}$ bands; 
$a_{\lambda}$: slope of the fit; 
$b_{\lambda}$: zero-point of the fit; 
rms: rms of the derived relations; 
$N$: number of stars used for fitting.}
\end{deluxetable*}

\begin{deluxetable*}{llcccc}
\tabletypesize{\scriptsize}
\tablewidth{0pt} 
\tablenum{3}
\tablecaption{The determined Period-Wesenheit relations for Galactic classical Cepheids \label{tab:pwr}}
\tablehead{
\colhead{Source} & \colhead{Wesenheit index} & \colhead{$a$} & \colhead{$b$} & 
\colhead{rms} & \colhead{N} 
}
\colnumbers
\startdata 
\multicolumn{5}{l}{Period-Wesenheit relation: $W = a(\rm{logP - 1}) + b$} \\ 
\hline
\hline 
{                 } & $W^{ri}_r = r - 4.051 (r-i) - \mu$ & -3.029 $\pm$ 0.097 & -4.645 $\pm$ 0.031 & 0.28 & 85 \\ 
{\citet{Green2019}} & $W^{gr}_r = r - 2.905 (g-r) - \mu$ & -3.511 $\pm$ 0.102 & -5.363 $\pm$ 0.032 & 0.29 & 85 \\ 
{                 } & $W^{gi}_g = g - 2.274 (g-i) - \mu$ & -3.240 $\pm$ 0.091 & -4.960 $\pm$ 0.029 & 0.26 & 85 \\ 
\hline 
{           } & $W^{ri}_r = r - 3.787 (r-i) - \mu$ & -2.934 $\pm$ 0.091 & -4.520 $\pm$ 0.029 & 0.26 & 85 \\ 
{\citet{F99}} & $W^{gr}_r = r - 2.385 (g-r) - \mu$ & -3.165 $\pm$ 0.081 & -4.891 $\pm$ 0.026 & 0.23 & 85 \\ 
{           } & $W^{gi}_g = g - 2.077 (g-i) - \mu$ & -2.963 $\pm$ 0.074 & -4.677 $\pm$ 0.023 & 0.21 & 84 \\ 
\hline 
\multicolumn{6}{l}{Astrometry-Based Luminosity: $ABL_{W} = 10^{0.2[a(logP - 1) + b]}$} \\ 
\hline
{                 } & $ABL_{W^{ri}_r} = \varpi 10^{(r-4.051(r-i) + 5)/5}$ & -2.829 $\pm$ 0.101 & -4.614 $\pm$ 0.030 & 0.29 & 85 \\ 
{\citet{Green2019}} & $ABL_{W^{gr}_r} = \varpi 10^{(r-2.905(g-r) + 5)/5}$ & -3.375 $\pm$ 0.108 & -5.338 $\pm$ 0.032 & 0.30 & 85 \\ 
{                 } & $ABL_{W^{gi}_g} = \varpi 10^{(g-2.274(g-i) + 5)/5}$ & -3.062 $\pm$ 0.088 & -4.933 $\pm$ 0.026 & 0.27 & 85 \\ 
\hline
{                 } & $ABL_{W^{ri}_r} = \varpi 10^{(r-3.787(r-i) + 5)/5}$ & -2.746 $\pm$ 0.097 & -4.491 $\pm$ 0.029 & 0.27 & 85 \\ 
{\citet{F99}      } & $ABL_{W^{gr}_r} = \varpi 10^{(r-2.385(g-r) + 5)/5}$ & -3.035 $\pm$ 0.088 & -4.871 $\pm$ 0.026 & 0.24 & 85 \\ 
{                 } & $ABL_{W^{gi}_g} = \varpi 10^{(g-2.077(g-i) + 5)/5}$ & -2.852 $\pm$ 0.081 & -4.659 $\pm$ 0.024 & 0.21 & 84 \\ 
\enddata
\tablecomments{ 
Source: source of extinction vector ($R_{\lambda}$); 
Wesenheit index: formula for calculation of the Wesenheit magnitude, where  
$\mu$ is the distance modulus; 
$a$: slope of the fit; 
$b$: zero-point of the fit; 
rms: rms of the derived relations; 
and $N$: number of stars used for fitting.
} 
\end{deluxetable*}


\begin{acknowledgments}
We thank the anonymous referee for valuable comments which improved this paper. 

The research leading to these results has received funding from the European 
Research Council (ERC) under the European Union’s Horizon 2020 research and 
innovation program (grant agreement No. 695099). We also acknowledge support 
from the National Science Center, Poland grants MAESTRO UMO-2017/26/A/ST9/00446, 
BEETHOVEN UMO-2018/31/G/ST9/03050, and DIR/WK/2018/09 grants of the Polish Ministry 
of Science and Higher Education. We also acknowledge financial support from 
UniverScale grant financed by the European Union’s Horizon 2020 research and
innovation programme under the grant agreement number 951549. We gratefully 
acknowledge financial support for this work from the BASAL Centro de
Astrofisica y Tecnologias Afines (CATA) AFB-170002 and the Millenium Institute 
of Astrophysics (MAS) of the Iniciativa Cientifica Milenio del Ministerio
de Economia, Fomento y Turismo de Chile, project IC120009. W.G. also gratefully 
acknowledges support from the ANID BASAL project ACE210002. 
P.W. gratefully acknowledges financial support from the Polish National 
Science Center grant PRELUDIUM 2018/31/N/ST9/02742. 
This work has made use of data from the European Space Agency (ESA) mission 
Gaia (\url{https://www.cosmos.esa.int/gaia}), processed by the Gaia Data Processing 
and Analysis Consortium (DPAC, \url{https://www.cosmos.esa.int/web/gaia/dpac/
consortium}). Funding for the DPAC has been provided by national institutions, 
in particular the institutions participating in the Gaia Multilateral Agreement.
\end{acknowledgments}

%

\vspace{5mm}
\facilities{LCOGT (0.4m)}


\software{{\tt gaiadr3\_zero-point} \citep{Lindegren2021b}, 
          IRAF \citep{Tody1986,Tody1993}, 
          DAOPHOT \citep{Stetson1987}, 
          Astropy \citep{Astropy2013},  
          Sklearn \citep{Pedregosa2011},
          NumPy \citep{Numpy1,Numpy2}, 
          SciPy \citep{Virtanen2020}, 
          and Matplotlib \citep{Hunter2007}.
          }

%
          


\appendix

\renewcommand\thefigure{\thesection.\arabic{figure}}

\section{The Sloan band light curves of Galactic classical Cepheids analazyed in this work.} 
\label{app:lc}

\setcounter{figure}{0}

Figure~\ref{fig:fig1} present the light curves of $96$ Galactic fundamental mode classical 
Cepheids used in this study. 

\begin{figure*}[h]
    \centering
    \includegraphics[width=0.9\textwidth]{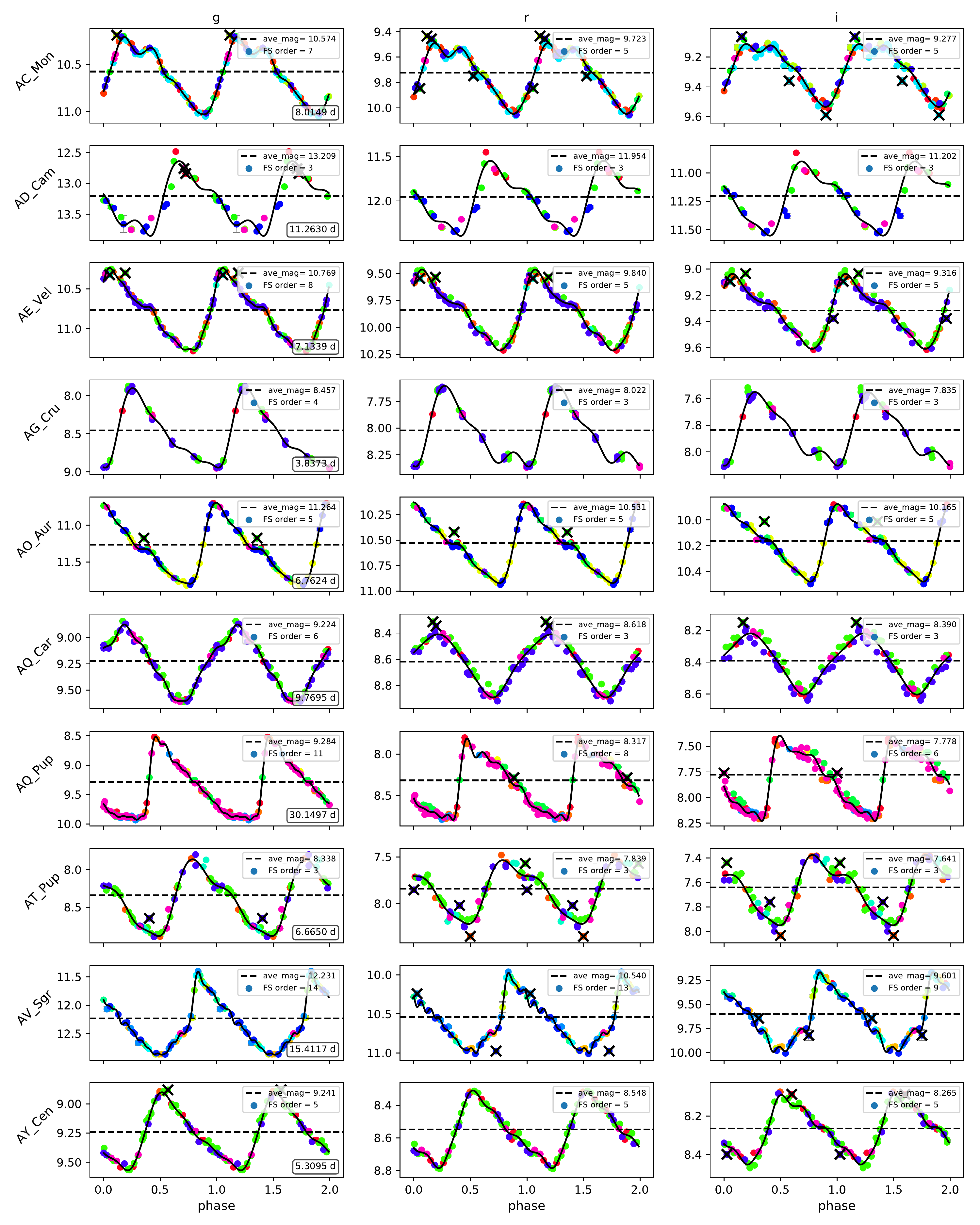}
    \caption{\label{fig:fig1} 
    The Sloan -- Pans-STARRS $g_{P1}r_{P1}i_{P1}$ band light curves of Cepheids analyzed 
    in this work. 
    Horizontal dashed, black lines correspond to the determined mean magnitudes. 
    Different colors of points mark different telescopes used during the data collection, 
    while black crosses mark points rejected from the fitting. 
    Black lines show the best fit Fourier series.}
\end{figure*}


\begin{figure*}
    \ContinuedFloat
    \centering
    \includegraphics[width=0.9\textwidth]{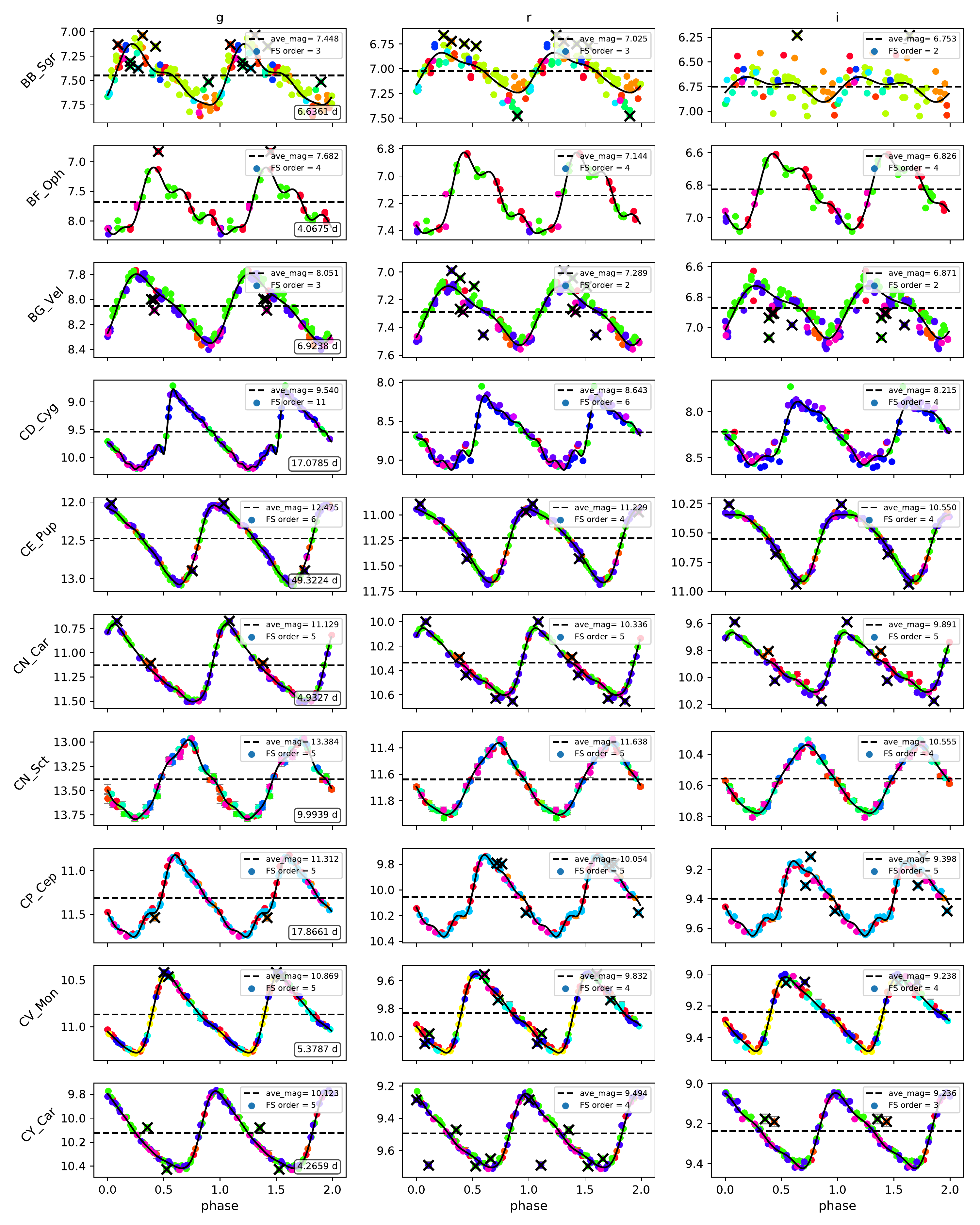}
    \caption{Continued from previous page.}
\end{figure*}

\begin{figure*}
    \ContinuedFloat
    \centering
    \includegraphics[width=0.9\textwidth]{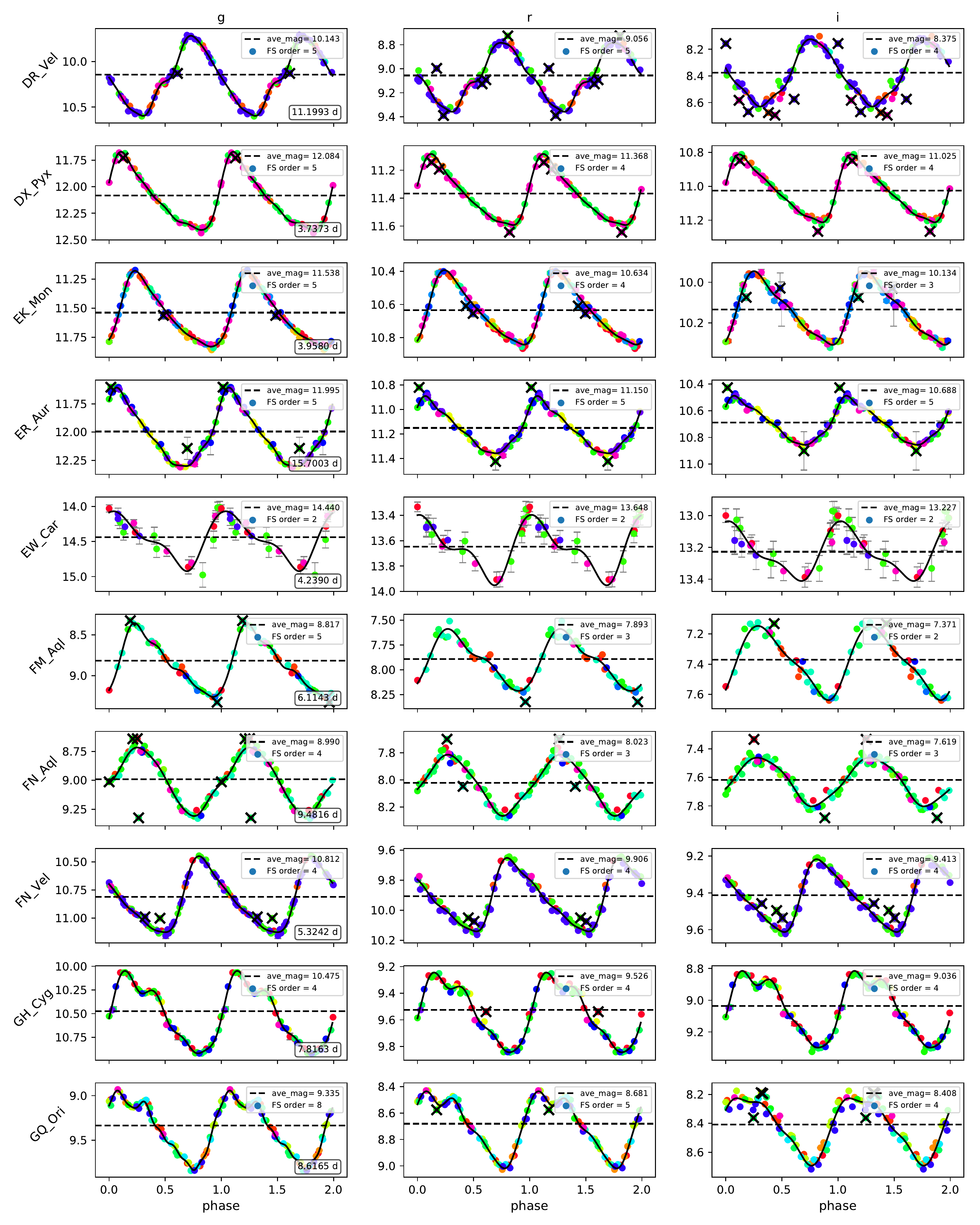}
    \caption{Continued from previous page.}
\end{figure*}

\begin{figure*}
    \ContinuedFloat
    \centering
    \includegraphics[width=0.9\textwidth]{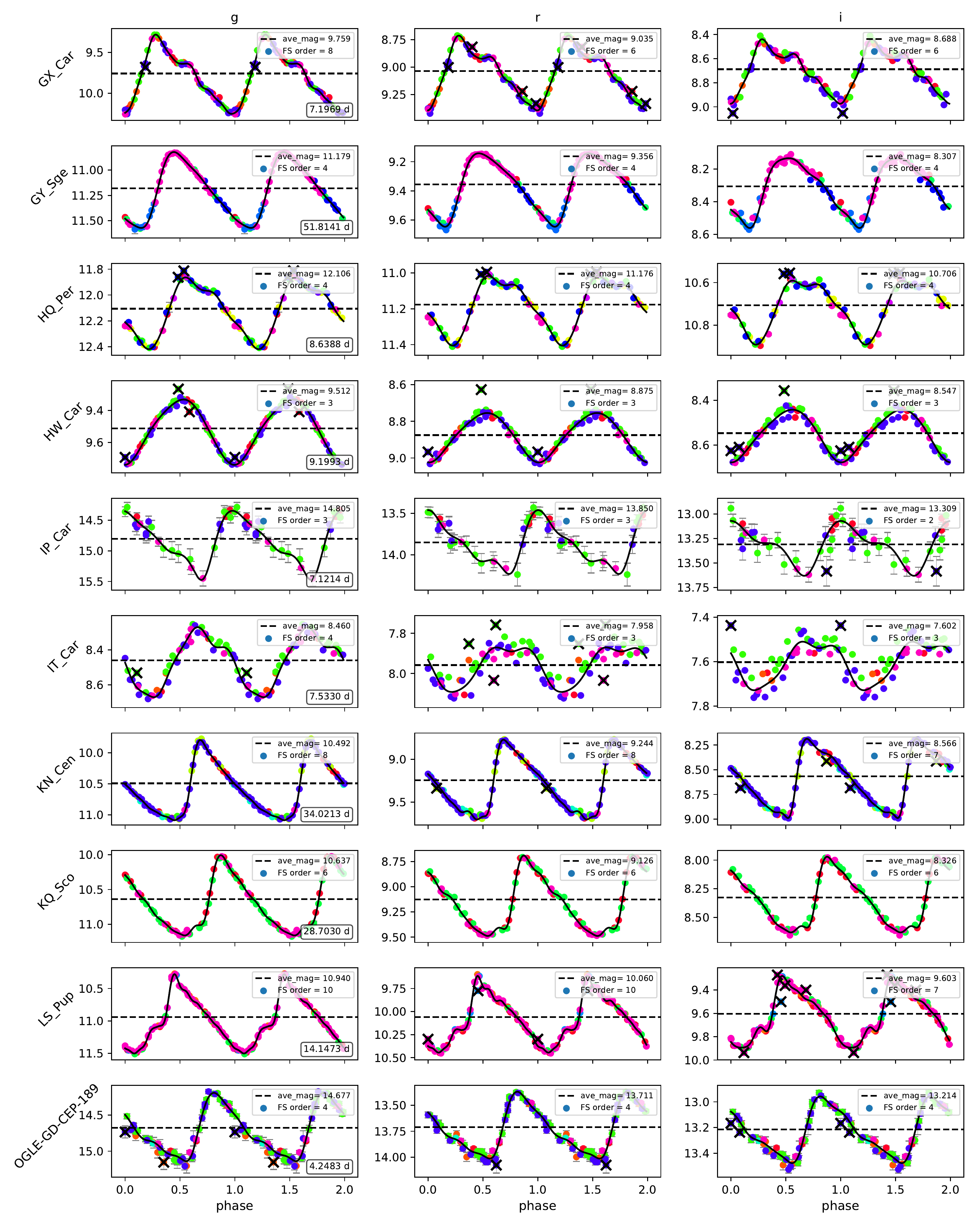}
    \caption{Continued from previous page.}
\end{figure*}

\begin{figure*}
    \ContinuedFloat
    \centering
    \includegraphics[width=0.9\textwidth]{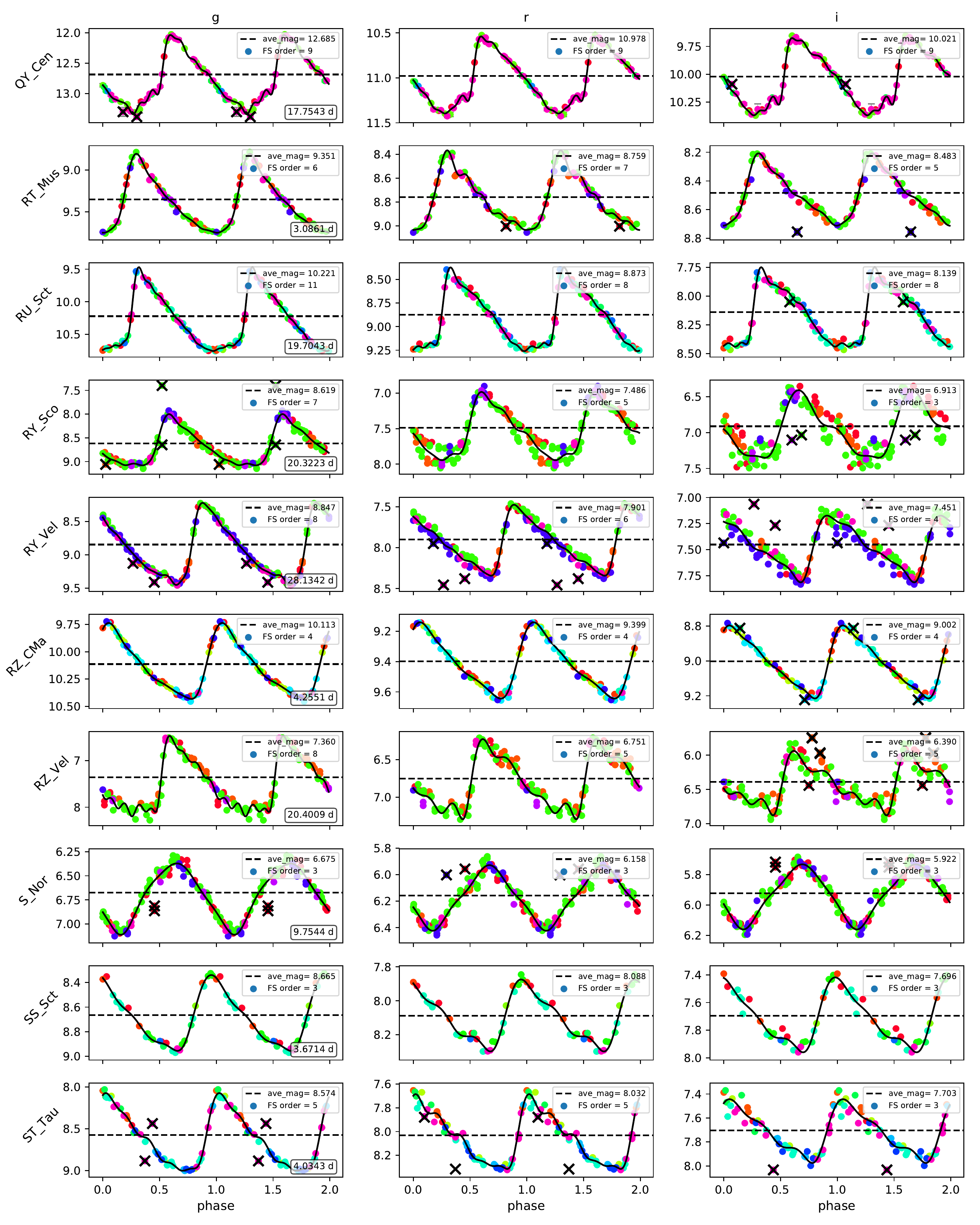}
    \caption{Continued from previous page.}
\end{figure*}

\begin{figure*}
    \ContinuedFloat
    \centering
    \includegraphics[width=0.9\textwidth]{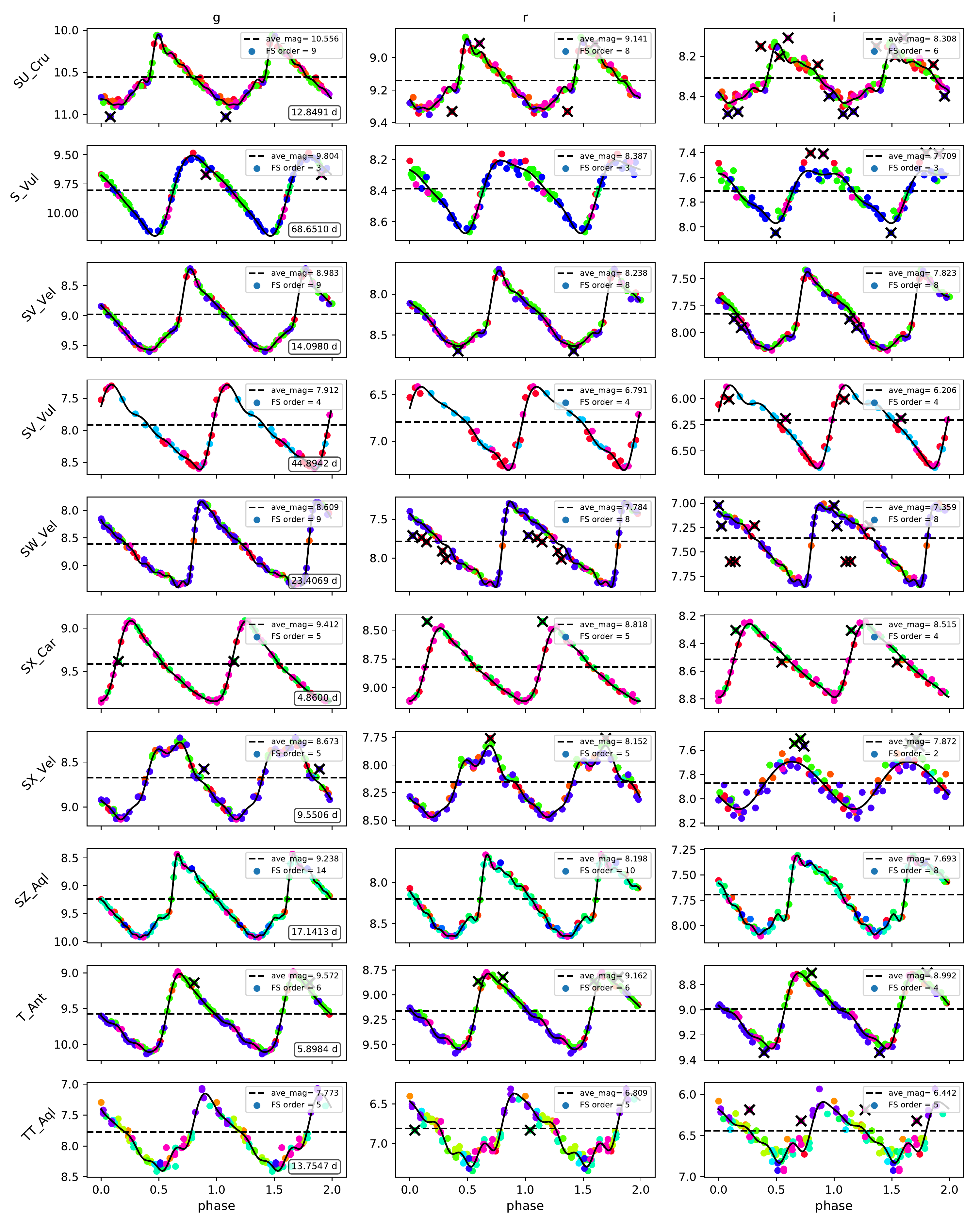}
    \caption{Continued from previous page.}
\end{figure*}

\begin{figure*}
    \ContinuedFloat
    \centering
    \includegraphics[width=0.9\textwidth]{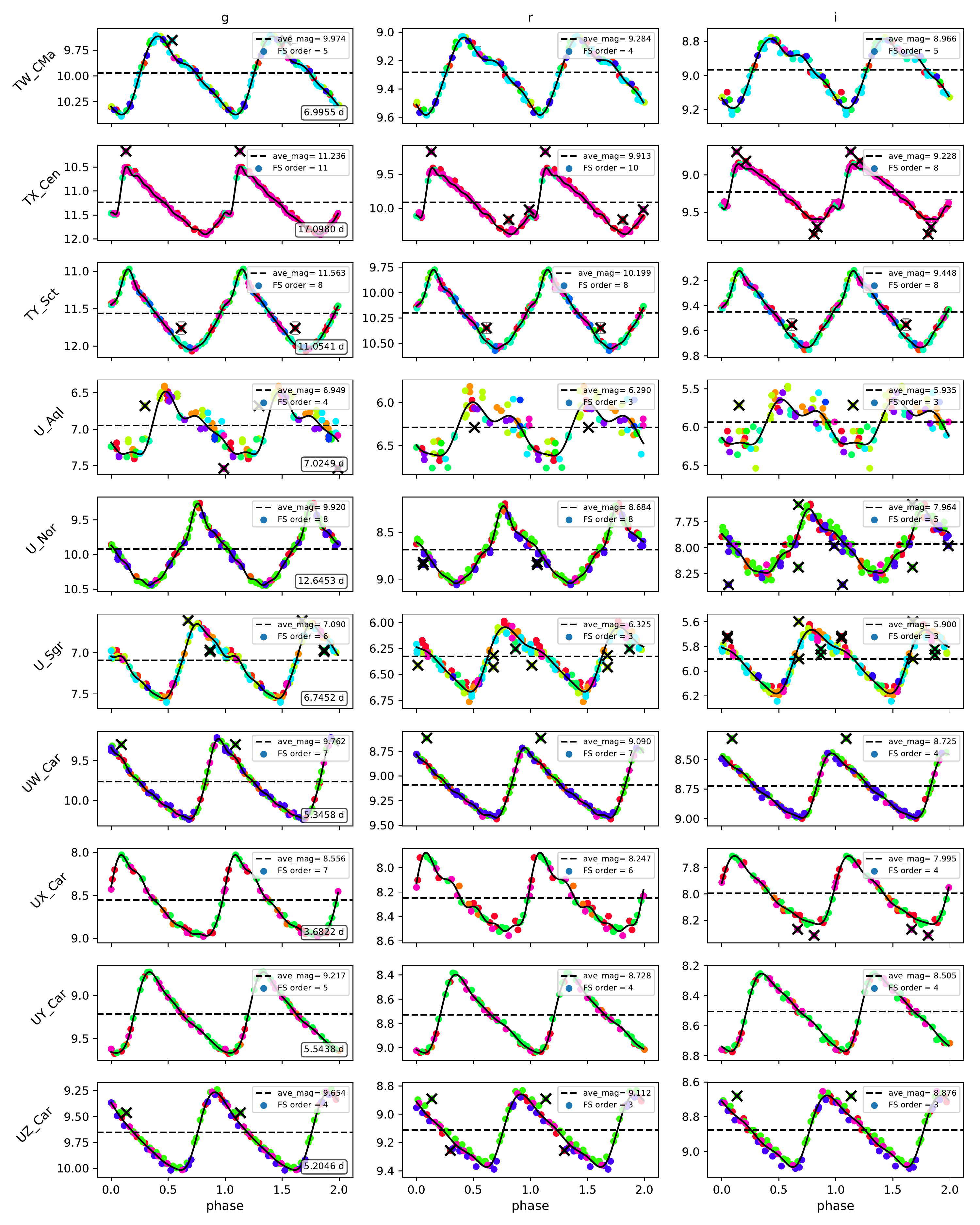}
    \caption{Continued from previous page.}
\end{figure*}

\begin{figure*}
    \ContinuedFloat
    \centering
    \includegraphics[width=0.9\textwidth]{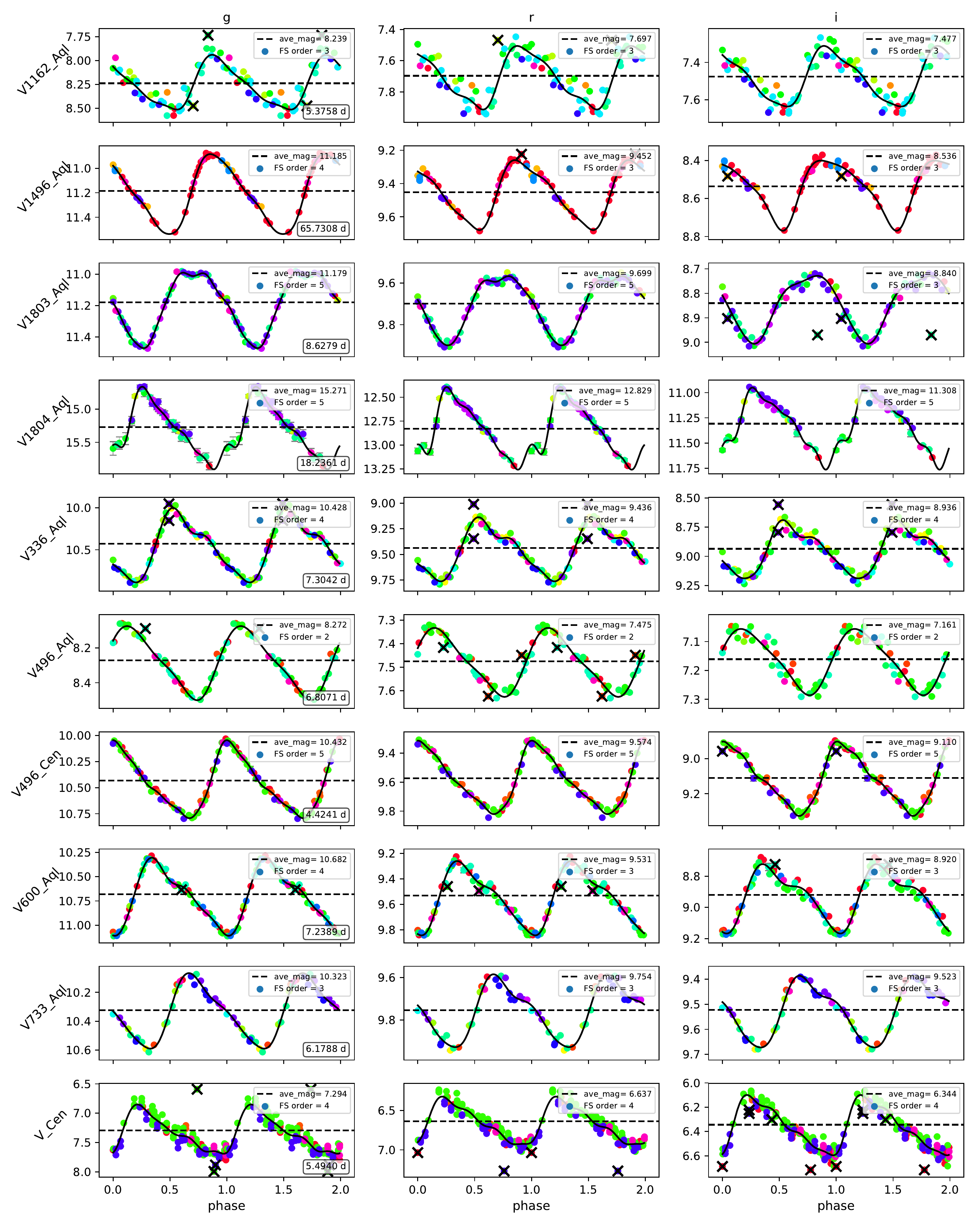}
    \caption{Continued from previous page.}
\end{figure*}

\begin{figure*}
    \ContinuedFloat
    \centering
    \includegraphics[width=0.9\textwidth]{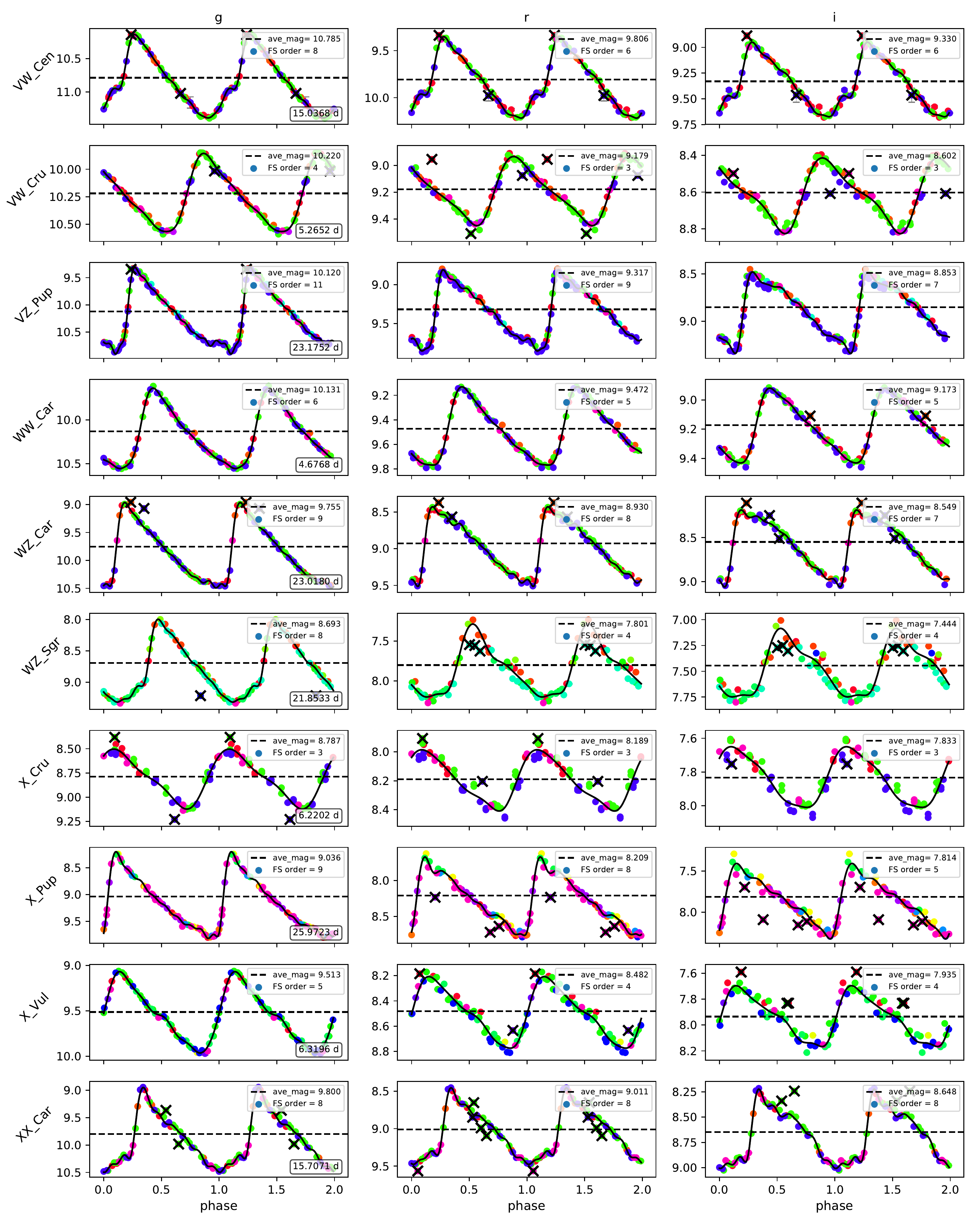}
    \caption{Continued from previous page.}
\end{figure*}

\begin{figure*}
    \ContinuedFloat
    \centering
    \includegraphics[width=0.9\textwidth]{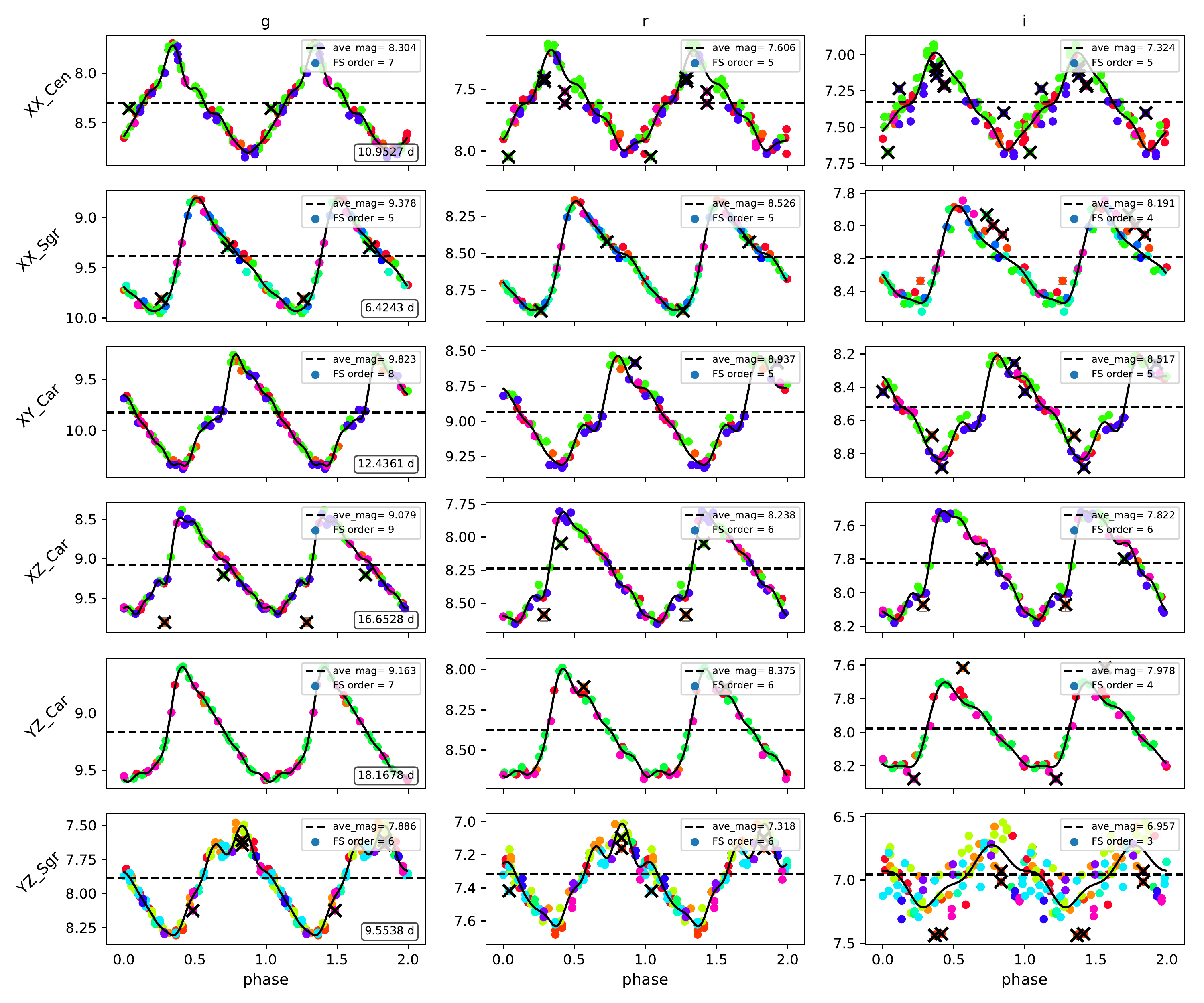}
    \caption{Continued from previous page.}
\end{figure*}



\begin{thebibliography}{} 
  \bibitem[Abazajian et al. (2003)]{Abazajian2003} Abazajian, K., Adelman-McCarthy, J.~K., Ag{\"u}eros, M.~A. et al. 2003, \aj, 126, 2081
  \bibitem[Adair \& Lee (2023)]{AL2023} Adair, S., Lee, C.-H. 2023, \aj, 165, 28 
  \bibitem[Anderson et al. (2016)]{Anderson2016} Anderson, R.~I., Saio, H., Ekström, S. et al. 2016, \aap, 591, 8 
  \bibitem[Arenou \& Luri (1999)] {AL1999} Arenou, F., Luri, X. 1999, ASPC, 167, 13 
  \bibitem[Astropy Collaboration (2013)]{Astropy2013}  Astropy Collaboration (Robitaille, T.~P., Tollerud, E.~J., Greenfield, P. et al.) 2013, \aap, 558, 33
  \bibitem[Bellm et al. (2018)]{Bellm2018} Bellm, E.~C., Kulkarni, S.~R., Graham, M.~J. et al. 2018, PASP, 131, 018002 
  \bibitem[Bellm et al. (2019)]{Bellm2019}  Bellm, E.~C., Kulkarni, S.~R., Barlow, T. et al. 2019, PASP, 131, 068003 
  \bibitem[Bono et al. (2010)]{Bono2010} Bono, G., Caputo, F., Marconi, M. et al. 2010, \apj, 715, 277 
  \bibitem[Breuval et al. (2021)] {Breuval2021} Breuval, L., Kervella, P., Wielg\'orski, P. et al. 2021, \apj, 913, 38
  \bibitem[Breuval et al. (2022)] {Breuval2022} Breuval, L., Riess, A.~G., Kervella, P. et al. 2022, \apj, 939, 89 
  \bibitem[Dark Energy Survey Collaboration et al. (2016)] {DES2016}  Dark Energy Survey Collaboration (Abbott, T. et al.) 2016, MNRAS, 460, 1270 
  \bibitem[Dekany et al. (2020)]{Dekany2020} Dekany, R., Smith, R.~M., Riddle, R. et al. 2020, PASP, 132, 038001 
  \bibitem[De Somma et al. (2022)]{DeSomma2022}  De Somma, G., Marconi, M., Molinaro, R. et al. 2022, ApJS, 262, 25 
  \bibitem[Di~Criscienzo et al. (2013)]{DiCriscienzo2013} Di~Criscienzo, M., Marconi, M., Musella, I. et al. 2013, \mnras, 428, 212 
  \bibitem[Feast \& Catchpole (1997)]{FC1997} Feast, M.~W., Catchpole, R.~M. 1997, MNRAS, 286, L1
  \bibitem[Fernie et al. (1995)] {Fernie1995} Fernie, J.D., Beattie, B., Evans, N.R., Seager, S. 1995, IBVS No. 4148  
  \bibitem[Fitzpatrick (1999)]{F99} Fitzpatrick, E.~L., PASP, 111, 63 
  \bibitem[Flaugher et al. (2015)]{Flaugher2015} Flaugher, B., Diehl, H.~T., Honscheid, K. et al. 2015, \aj, 150, 150 
  \bibitem[Fouqu{\'e} et al. (2007)]{Fouque2007} Fouqu{\'e}, P., Arriagada, P., Storm, J. et al. 2007, \aap, 476, 73 
  \bibitem[Fukugita et al. (1996)]{Fukugita1996} Fukugita, M., Ichikawa, T., Gunn, J.~E. et al. 1996, \aj, 111, 1748 
  \bibitem[Gaia Collaboration (2023)]{GaiaCol2023} Gaia Collaboration (Vallenari, A. et al.) 2023, \aap, 674, A1 
  \bibitem[Gieren et al. (2018)]{Gieren2018} Gieren, W., Storm, J., Konorski, P. et al. 2018, \aap, 620, A99 
  \bibitem[Green et al. (2019)] {Green2019} Green, G.~M., Schlafly, E., Zucker, C. et al. 2019, \apj, 887, 93
  \bibitem[Groenewegen (2018)]{Groenewegen2018} Groenewegen, M. A. T. 2018, \aap, 619, A8 
  \bibitem[Groenewegen (2021)]{Groenewegen2021} Groenewegen, M. A. T. 2021, \aap, 654, A20 
  \bibitem[Harris et al. (2020)]{Numpy2} Harris, C.~R., Millman, K.~J., van der Walt, S.~J., et al. 2020, Natur, 585, 357 
  \bibitem[Harris (1981)]{Harris1981} Harris, H.~C. 1981, \aj, 86, 1192 
  \bibitem[Hoffmann \& Macri (2015)]{HM2015} Hoffmann, S.~L., Macri, L.~M. 2015, \aj, 149, 183 
  \bibitem[Hunter et al. (2007)]{Hunter2007} Hunter, J.~D. 2007, CSE, 9, 90 
  \bibitem[Ivezi\'c et al. (2019)]{Ivezic2019} Ivezi\'c, \v{Z}, Kahn, S.~M., Tyson, J.~A., et al. 2019, \apj, 873, 111
  \bibitem[Karczmarek et al. (2023)]{Karczmarek2023} Karczmarek, P., Hajdu, G., Pietrzyński, G. et al. 2023, arXiv:2303.15664
  \bibitem[Kodric et al. (2018)]{Kodric2018} Kodric, M., Riffeser, A., Hopp, U. et al. 2018, \aj, 156, 130 
  \bibitem[Kovtyukh et al. (2008)]{Kovtyukh2008} Kovtyukh, V.~V., Soubiran, C., Luck R.~E. 2008, \mnras, 389, 1336 
  \bibitem[Lallement et al. (2019)] {Lallement2019} Lallement, R., Babusiaux, C., Vergely, J.~L. et al. 2019, \aap, 625, 135
  \bibitem[Leavitt (1908)] {Leavitt1908} Leavitt, H. S. 1908, AnHar, 60, 87
  \bibitem[Leavitt \& Pickering (1912)]{LP1912} Leavitt, H. S., Pickering, E. C. 1912, HarCi, 173, 1 
  \bibitem[Li et al. (2021)]{Li2021} Li, S., Riess, A.~G., Busch, M.~P. et al. 2021, \apj, 920, 84 
  \bibitem[Lindegren et al. (2021a)] {Lindegren2021a} Lindegren, L., Klioner, S.~A., Hern\'andez, J. et al. 2021a, \aap, 649, A2
  \bibitem[Lindegren et al. (2021b)] {Lindegren2021b} Lindegren, L., Bastian, U., Biermann, M. et al. 2021b, \aap, 649, A4 
  \bibitem[Macri et al. (2015)] {Macri2015} Macri, L. M., Ngeow, C.-C., Kanbur, S. M. et al. 2015, AJ, 149, 117 
  \bibitem[Madore (1982)]{Madore1982} Madore, B.~F. 1982, \apj, 253, 575  
  \bibitem[Madore, Freedman \& Moak (2017)]{MFM2017} Madore, B.~F., Freedman, W.~L., Moak, S. 2017, \apj, 842, 42 
  \bibitem[Molinaro et al. (2023)]{Molinaro2023} Molinaro, R., Ripepi, V., Marconi, M. et al. 2023, \mnras, 520, 4154 
  \bibitem[Ngeow et al. (2022a)] {Ngeow2022_RRL} Ngeow, C.-C., Bhardwaj, A., Dekany, R. et al. 2022a, \aj, 163, 239 
  \bibitem[Ngeow et al. (2022b)] {Ngeow2022_T2CEP} Ngeow, C.-C., Bhardwaj, A., Henderson, J.~Y. et al. 2022b, \aj, 164, 154
  \bibitem[Ngeow et al. (2022c)] {Ngeow2022_AC} Ngeow, C.-C., Bhardwaj, A., Graham, M.~J. et al. 2022c, \aj, 164, 191 
  \bibitem[Ngeow et al. (2023)]{Ngeow2023_SXPhe} Ngeow, C.-C., Bhardwaj, A., Graham, M.~J. et al. 2023, \aj, 165, 190  
  \bibitem[Pedregosa et al. (2011)] {Pedregosa2011} Pedregosa, F., Varoquaux, G., Gramfort, A. et al. 2011, Journal of machine learning research, 12, 2825  
  \bibitem[Pietrukowicz, Soszy\'nski \& Udalski (2021)] {PSU2021} Pietrukowicz, P., Soszy\'nski, I., Udalski, A. 2021, AcA, 71, 205
  \bibitem[Riess et al. (2021)]{Riess2021} Riess, A.~G., Casertano, S., Yuan, W., et al. 2021, ApJL, 908, L6  
  \bibitem[Riess et al. (2022)] {Riess2022}  Riess, A~G., Yuan, W., Macri, Lucas M. et al. 2022, \apj, 934, 7 
  \bibitem[Ripepi et al. (2016)]{Ripepi2016} Ripepi, V., Marconi, M., Moretti, M.~I. et al. 2016, ApJS, 224, 21 
  \bibitem[Ripepi et al. (2020)]{Ripepi2020} Ripepi, V., Catanzaro, G., Molinaro, R. et al. 2020, \aap, 642, A230 
  \bibitem[Schlegel, Finkbeiner \& Davis (1998)] {SFD} Schlegel, D.~J., Finkbeiner, D.~P., Davis, M. 1998, \apj, 500, 525  
  \bibitem[Smith et al. (2002)]{Smith2002}  Smith, J.~A., Tucker, D.~L., Kent, S. et al. 2002, \aj, 123, 2121 
  \bibitem[Stetson (1987)] {Stetson1987} Stetson, P. B. 1987, PASP, 99, 191  
  \bibitem[Szabados (1996)]{Szabados1996} Szabados, L. 1996, \aap, 311, 189 
  \bibitem[Tammann et al. (2003)]{Tammann2003} Tammann, G.~A., Sandage, A., Reindl, B. 2003, \aap, 404, 423 
  \bibitem[Tody (1986)]{Tody1986} Tody, D. 1986, Proc. SPIE, 627, 733
  \bibitem[Tody (1993)]{Tody1993} Tody, D. 1993, in ASP Conf. Ser. 52, Astronomical Data Analysis Software and Systems II, ed. R. J. Hanisch, R. J. V. Brissenden, \& J. Barnes (San Francisco, CA: ASP), 173 
  \bibitem[Tonry et al. (2012)]{Tonry2012} Tonry, J.~L., Stubbs, C.~W., Lykke, K.~R. et al. 2012, \apj, 750, 99 
  \bibitem[Tonry et al. (2018)] {Tonry2018} Tonry, J.~L., Denneau, L.,  Flewelling, H. et al. 2018, \apj, 867, 105 
  \bibitem[Trentin et al. (2023)]{Trentin2023} Trentin, E., Ripepi, V., Catanzaro, G. et al. 2023, \mnras, 519, 2331 
  \bibitem[Turner (1989)]{Turner1989} Turner, D.~G. 1989, \aj, 98, 2300 
  \bibitem[Turner (2016)]{Turner2016} Turner, D.~G. 2016, RMxAA, 52, 223 
  \bibitem[Udalski et al. (1999)]{Udalski1999} Udalski, A., Szymanski, M., Kubiak, M. et al. 1999, AcA, 49, 201 
  \bibitem[van der Walt et al. (2011)]{Numpy1} van der Walt, S., Colbert, S. C., Varoquaux, G. 2011, CSE, 13, 22
  \bibitem[Virtanen et al. (2020)]{Virtanen2020} Virtanen, P., Gommers, R., Oliphant, T. E., et al. 2020, NatMe, 17, 261 
  \bibitem[Wang et al. (2018)]{Wang2018}  Wang, S., Chen, X., de Grijs, R. et al. 2018, \apj, 852, 78 
  \bibitem[Wielg\'orski et al. (2022)]{Wielgorski2022} Wielg\'orski, P., Pietrzy\'nski, G., Pilecki, B. et al. 2022, \apj, 927, 89
\end{thebibliography}



\end{document}